.tfm scaled 1000 %  BGotyk
 \font\eufm=eufm10.tfm scaled 1000 %  Gotyk
 \font\rsfs=rsfs10.tfm scaled 1000 %  Kaligrafik
\documentclass[12pt,a4paper,oneside]{article}
\usepackage[psamsfonts]{amssymb}
\usepackage{amsmath}
\usepackage{amscd}
\usepackage{ifthen}
\def\car#1#2{[   \raisebox{-.6mm}{$\vec{  #1} #2$} ]}
\def\carr#1#2#3{[  #3 \raisebox{-.6mm} {$\vec{  #1\,}#2$} ]}
\def\c #1{\mbox{\rsfs  #1}} %  kalig

\def\bo#1{\boldsymbol{#1}}
\def\vth{\vartheta}

\def\g #1{\mbox{\eufm  #1}} %  Gotyk
\def\Re{\mathbb{R}}
\def\Ce{\mathbb{C}}

\def\tsc#1{\textsc{#1}} \def\tsf#1{{\mathsf{ #1}}}

\def\sign{{\rm sign}}
\def\ds{\displaystyle}    
\def\nonu{\nonumber}
               \def\noi{\noindent}
  \def\pl{\partial}
\def\Gam{\Gamma}          \def\gam{\gamma}
\def\vp{\varphi}          
\def\eps{\epsilon}        \def\lam{\lambda}
\def\Lam{\Lambda}         \def\om{\omega}        \def\Om{\Omega}
\def\eps{\epsilon}        
\def\a{\alpha}            \def\bt{\beta}
\def\Del{\Delta}          \def\del{\delta}
\def\kap{\kappa}
\def\co{{{}^\ast}}        \def\cov{{{}_\ast}}
\def\Sig{\Sigma}          \def\we{\wedge}
\def\sig{\sigma}
\def\sS#1{\!#1\!}         
\def\ra{\rangle}          \def\la{\langle}      
\def\ord#1{{}^{(#1)}}

\def\trx#1{ {^{\rm t \mspace{#1mu} }}}

\def\hph#1{{\hphantom{#1}}}      \def\vph#1{{\vphantom{#1}}}

\def\llist#1{ \begin{list}{}{\itemindent=2mm\parsep=0mm\itemsep=0mm\topsep =1mm} #1 \end{list} }
\def\itt#1{ \item[$(#1)$]  }

\def\pdown#1#2#3{  \mathop{ #1} \llap{ \raise #3     \hbox{$\ssssty \hfill #2 \hfill $}}}  %       put down
\def\lamp{\lambda\negthinspace\llap{\raise.25cm\hbox{$\ssssty\circ$} }}
\def\rhoo{\rho \llap{ \raise.15cm\hbox{$\ssssty \circ $}}}

\def\lzg#1#2{\stackrel{\raise #2 \hbox{$\scriptscriptstyle\circ $}} {#1}}

\def\con{{}_-{\raise .054cm \hbox{$\ssssty \hskip-.088cm \mid$}}}
\def\gb#1{\setbox0=\hbox{$#1$} \kern-.025em\copy0\kern-\wd0\kern.025em
\copy0\kern-\wd0 \kern-.025em\raise.0333em\box0}
\def\pr{^\prime}  

\def\xmm#1#2#3#4{ #3\beg{smallmatrix}{#1\\#2} #4}
\def\beq{\begin{equation}}
\def\eeq{\end{equation}}
\def\beg#1#2{ \begin{#1}   #2 \end{#1}}
\def\beqm#1{ \beg{equation} {#1} }
\def\begg#1{ \begin{gather} #1 \end{gather} }

\def\lab{\label}

\def\Meq#1#2{\begin{subequations} \lab{#1} {#2}  \end{subequations}}
\newtheorem{theo}{{\em Theorem}}
\newtheorem{defin}{{\em Definition}}
\def\ndef#1{\begin{defin}$\!\!\!$.#1 \end{defin}}
\def\ntheo#1{\begin{theo}$\!\!\!$.#1 \end{theo}}

\usepackage{setspace}
\usepackage{verbatim}
\usepackage[cp1250]{inputenc}
\usepackage{epsfig} %\usepackage{changebar}
\usepackage{graphics}

\def\geo#1#2{ {\la \vec #1 \ra _g}{^{#2} } }

\def\toast{ \vph{|}_\ast}
\def\toastx#1{\vph{|}_{#1}}
\def\xmm#1#2#3#4{ #3\beg{smallmatrix}{#1\\#2} #4}
\def\tsc#1{\textsc{#1}}
\def\vth{\vartheta}
\def\vk{\varkappa}
\begin{titlepage}
\title{Ellipsoids of ${\rm U}(3)$ model}
\author{M. Cerkaski$^{1,2,}$\footnote{ Electronic Mail: cerkaski@thsun1.jinr.ru }}
\end{titlepage}
\begin{document}
\makeatletter
\renewcommand\@biblabel[1]{\textsuperscript{#1}}
\renewcommand{\@cite}[2]{{\nolinebreak[4]{\textsuperscript{#1}}\ifthenelse{\boolean{@tempswa}}{,#2}{} }}
\makeatother
\maketitle
\renewcommand{\thesection}{\Roman{section}}
\beg{center}{ \noi${}^1$ Henryk Niewodnicza\'{n}ski Institute of Nuclear Physics PAN,
 Department of Theoretical Physics, ul. Radzikowskiego 152, 31--342,  Krak\'{o}w, Poland }
% end{center}
\beg{center}{${}^2$ Bogoliubov Laboratory of Theoretical Physics,
         Joint Institute for Nuclear Research,  141980 Dubna, Russia }
%
% $^b$ Bogoliubov Laboratory of Theoretical Physics, JINR,
%   Dubna, Russia and CSNSM-IN2P3, Orsay, France \\[2mm]\vskip 1.cm
%
\begin{abstract}
The Cartan model of ${\rm SO}(3)/{\rm SO}(2)$ matrices
is applied to  reduce of rotational degrees of freedom on coadjoint orbits
of ${\rm u}^\ast(3)$  Poisson algebra.
The seven--dimensional Poisson algebra  ${\rm u}_{{\rm SO}}(3)$   obtained by
${\rm SO}(3)$ reduction of ${\rm u}^\ast (3)$ algebra is found and
canonical parametrization of ${\rm u}^\ast(3)$ orbits $[p_1,p_2,p_3]\toast$ is studied.

The structure of bands formed by so--called families of $\tsf{S}$ and $\tsf{P}$ ellipsoids obtained
by searching  extremes of many--body  ${\rm SO}(3)$ invariant Hamiltonians is investigated.

The reduced four--dimensional system of equations of motion  describing the
simple schematic Hamiltonian based on the volume conservation is presented.
A new set of canonical coordinates regarding   the separation of
motion for independent modes is found  with the help of the Jacobi approach.
Bohr Somerfield's quantization of new momentum space is studied. \end{abstract}
%
%\documentclass[12pt,a4paper,oneside]{article}      %\usepackage{amssymb,amsmath}
%\href{}{}
%
\vfill
\eject
\renewcommand{\baselinestretch}{1.5}
\section{Introduction}
The aim of this paper is to discuss the simplest analytically solvable model
describing the separation of collective degrees of freedom of many--particle dynamics
into the pure rotational modes and  the intrinsic one represented by ${\rm SO}(3)$ scalar functions.

The main role of the approach based on discussion of coadjoint orbits $[p_1,p_2,p_3]\toast$ of ${\rm u}(3)$ algebra
is to receive a preliminary  description of the structure of  equilibrium figures occurring in a more advanced and
bristling with technical difficulties model
based on coadjoint orbits $\la p_1,p_2,p_3\ra$ of ${\rm Sp}(6,R)$ group\cite{cerkJMP}
and apply earlier studies\cite{RR2,RR3,Rowe,cerkAn} employing the ${\rm Sp}(6,R)\times{\rm SO}(N-1)$ group for construction
of a curvilinear system of coordinates on N--body phase space.
The presented simplified model  provides  valuable results for explanation of some
general features of  the nuclear collective spectra of the ${\rm Sp}(6,R)$ model
if the solutions describe  almost  spherical symmetric  systems.
The latter are selected forcing the conditions: $p_1-p_2\ll p_3,p_2-p_3\ll p_3$.

The structure of Hamiltonian extremes
on $[p_1,p_2,p_3]\toast$ orbits  is  explained by considering two families of bands.

The first family of bands contains the equilibrium   figures called   $\tsf{S}$ ellipsoids,
while   the second one contains $\tsf{P}$ ellipsoids.
Here $\tsf{S}$ and $\tsf{P}$ ellipsoids generalize    the concept of $\tsf{S}$  and $\tsf{P}$ ellipsoids
introduced by Riemann for classification of equilibrium figures of   the
Dirichlet, Dedekind, Riemann model\cite{Chandra}.

In the paper we present the construction of a system of canonical coordinates on coadjoint orbits of ${\rm u}(3)$
regarding the reduction of dynamics of the ${\rm SO}(3)$ invariant system.

The first section  employs elements  of Cartan  ${\rm SO}(3)/{\rm SO}(2)$  transformations  in order to get
\begin{list}{}{\itemindent=2mm\parsep=0mm\itemsep=0mm\topsep =0mm}
\item[$(a)$]  ${\rm SO}(2)$ reduction of many particle phase space,
\item[$(b)$]  Poisson bracket $\{.,.\}\toast$   on  ${\rm SO}(2)$ reduced space,
\item[$(c)$]  reduction of ${\rm u}\co(3)$ to ${\rm u}_{\rm SO}(3)$.
\end{list}
In section III, four dimensional phase spaces $M^\eps_{\vec{p}}$ are  applied
to  construct  the canonical parametrization of $[p_1,p_2,p_3]\toast$ orbits.
In section IV, families of $\tsf{P}$ and $\tsf{S}$ ellipsoids are discussed.
In section V, using  a simple  class of ${\rm SO}(3)$--scalar ${\rm u}\co(3)$ Hamiltonians
and applying the Jacobi approach we obtain a new canonical parametrization  $[p_1,p_2,p_3]\toast$.
In the last section,  the quantum spectrum of the pair of new canonical momenta
is searched with the help of  Bohr Somerfield's  rules of  quantization.
Formulas determining the rules associated with point $(b)$ are found in Appendix.

\section{ ${\rm SO}(3)$ reduced functions  on orbits of $u\co(3)$ algebra }
%
%\section{ ${\rm O}(3)$ reduction of $u\co(3)$ to $u^\ast_{{\rm SO}(3)}(3)$  Poisson algebra. }
%
In the  application of group transformation $(\tsf{\tsf{G}},\tsf{M})$ to Hamiltonian dynamics,
elements of Lie algebra  $U_t \sS= \dot g_t\cdot  {g_t}{^{-1}}\in \g g$  are  obtained
studying the mapping $t\to m_t = g_t\bo{\cdot} m_0$ under  the assumption
$g_1\cdot (g_2\bo \cdot m)\sS=(g_1\cdot g_2)\bo{\cdot} m$ and considering the following formula:
\beqm{ \dot m_t \sS= \dot g_t \bo{\cdot}  m_0 =(\tsf{U}_t \cdot g_t) \bo \cdot  m_0\sS= \tsf{U}_t\bo \cdot m_t. }
where $\dot x \sS\equiv d x/d t $. In particular,  in the case   $\tsf{G}\sS= {\rm U}(3)$,  $g_t$
are complex three dimensional matrices $g_t\cdot {g_t}{^\dag} \sS= \bo 1$ for which
Lie algebra is spanned  by elements: $\g g\sS=u(3) \sS= \{\tsf{A}  =
\Sig_{\a \bt}\,\tsf{E}_{ab}\, A_{\a\bt},\, A_{ab} \sS=-A_{ba}\co \}$ where
$\tsf{E}_{ab}$ denote 3-dimensional matrices the  elements of which  read: $(\tsf{E}_{ab})_{cd}\sS=\del_{ac}\,\del_{bd}$.

Within application to the particle dynamics, ${\rm U}(3)$ group follows from  the study of group chain reduction
${\rm U}(3) \subset {\rm Sp}(6,R) \subset {\rm Sp}(6\,N,R)$
 where ${\rm Sp}(6\,N,R)$ is the group of linear canonical transformations of $N$--particle space,
 while ${\rm Sp}(6,R)$  span the subspace of  the collective one.
%
%\eject
%
Let  $m \equiv \binom{\bo a^1}{\bo a^2},\,\bo a^u=(\vec a^u_1,\ldots \vec a^u_N),\,u\sS=1,2$  be
a parametrization of points in $6\times N$  dimensional phase space.
The application of ${\rm U}(3)$ group to the particle dynamics bases on the following
formulas:
\begin{list}{}{\itemindent=2mm\parsep=0mm\itemsep=0mm\topsep =2mm}
\item[$(a)$]  $a^u_{a n}\sS= ( 2\,\kap_{a n}\,\hbar\,)^{-1/2}\, [ p_{a n} + (-1)^u\,i\,\kap_{a n} \,r_{an}]\sS=
     (a^{3-u}_{an})\co,\; a=x,y,z,\,n=1,\ldots,N$,
\item[$(b)$] $\Om(p,r)= \Sig_{in}\,p_{in}\we r_{in} \Leftrightarrow \{ a^u_{an}, a^v_{am} \}
       \vph{|} = -i\,(-1)^{u-1}\,\del_{u\,3-v}\,\del_{ab}\,\del_{nm}$,
\item[$(c)$] $A_{ab}(m) = (a^2 \cdot {^{\rm t\,}}a^1)_{ab} -  (\kap^{-1}\,P_a+ i\,\kap\,X_a)\, (\kap^{-1}\,P_b-  i\,\kap\,X_b)$,
\item[$(d)$]  $\tsf{A} (m) =
    \Sig_{ab}\, A_{ab}(m)\,\tsf{E}_{ab}\sS=\tsf{A} ^\dag(m)$,
\item[$(e)$]  $A_{ab}(m) =   {\rm trace}\, \tsf{A} (m)\cdot \tsf{E}_{ba} $,
\item[$(f)$] $\tsf{A}_t=\tsf{A} (m_t)$, $m_t=\xmm{{g_t}\co  & \bo 0 }{\bo 0 &g_t}{\bigl(}{\bigr)}\cdot m_0$ then
         $\tsf{A}_t = g_t \cdot \tsf{A}_0 \cdot g^{-1}_t = {\rm Ad}_{g_t}(\tsf{A}_0)$ and   $i\,\tsf{A}(m)\in {\rm u}(3)$,
\item[$(g)$] $\dot{\tsf{A}}_t = [\tsf{U}_t,\tsf{A}_t]$,  $\dot x_t \equiv \tfrac{d~}{dt} x_t$,
\item[$(h)$] $\tsf{A}(m)= \sum_\mu Z^\mu(m)\; \trx {1}  \tsf{Z}_{\mu}=
  \sum_{\mu \nu} G_{\mu\nu}[\tsf{Z}]\, Z_\mu(m)\,\trx 1 \tsf{Z}_{\mu}$,
\item[$(i)$] $\{Z_\mu,Z_\nu\}_m = -i\, {\rm trace}\,\tsf{A}(m)\cdot {^{\rm t}}\,[\tsf{Z}_\mu,\tsf{Z}_\nu]$,
\end{list}
where   $\kap_{a n}= \om_a\,\tsf{m}_n$, $\kap_n =\om\,\tsf{m}_n$,  $\om_x\sS=\om_y\sS= \om_z\sS= \om$,
$(\tsf{m}_1,\ldots, \tsf{m}_{N})$ are  the particle masses, $Z_\mu(m)$ and $\tsf{Z}_\mu$
are components of vectors $\vec Z(m) \sS=(Z_1(m),\ldots,Z_9(m))$
and the vector of matrices   $[\tsf{Z}]\sS=[\tsf{Z}_1,\ldots,\tsf{Z}_9]$, respectively.
We chose them  using two pairs of bases: %
$(\vec Z_\Ce(m),[\tsf{Z} _\Ce])$ or
$(\vec Z_\Re(m), [\tsf{Z}_\Re])$ where
$\vec Z_\Re(m) \sS=\lam(\vec Z_\Ce(m))$ and   $[\tsf{Z}_\Re]\sS=\kap([\tsf{Z}_\Ce])$. Here,
\begin{list}{}{\itemindent=2mm\parsep=0mm\itemsep=0mm\topsep =1.5mm}
\item[$(j)$]  $\kap$: \, $\tsf{L} _a = -i\,\Sig_{bc}\,\eps_{abc}\,\tsf{E}_{bc}$,  $\tsf{Q}_{ab}  =
   \tfrac 1 2\,(\tsf{E}_{ab} + \tsf{E}_{ba})$,
\item[$(k)$] $\lam$: \, $\tsf{L}_a(m) = -i\,\Sig_{bc}\,\eps_{abc}\,A_{bc}(m),\, \tsf{Q}_{ab}(m) =
             \tfrac 1 2\,(A_{ab}(m)+ A_{ba}(m))$,
\item[$(l)$]   $\kap\ord{-1}$: \, $\tsf{E}_{ab} =  \tsf{Q}_{ab} + \tfrac i 2\,\Sig_{k}\,\eps_{abc}\,\tsf{L}_c$,\;
             $\lam\ord{-1}$: \,  $A_{ab}(m) = \tsf{Q}_{ab}(m) + \tfrac i 2\,\Sig_{c}\,\eps_{abc}\,\tsf{L}_c(m)$,
\end{list}
hence
\begin{list}{}{\itemindent=2mm\parsep=0mm\itemsep=0mm\topsep =1.5mm}
 \item[$(m)$] $\tsf{L}_a(m)\sS =  \Sig_{bc}\, \eps_{abc}\,[ \Sig_{n} x_{an}\,p_{bn} - X_{a}\,P_b]   = 2\,\tsf{L}^a(m)\in \Re$,
\item[$(n)$]  $\tsf{Q}_{ab}(m)= \tfrac 1 2\,[ \Sig_{n}(\kap_{n}\,x_{an}\,x_{bn}+ {\kap_{n}}{^{-1}} \,p_{an}\,p_{bn}) -
     \kap\,{X_a}{^2} - \kap^{-1}\,{P_a}{^2}    =(2-\del_{ab})^{-1}\,\tsf{Q}^{ab}(m)\in \Re$,
\item[$(o)$] $\tsf{A}(m) =  \sum_a \tsf{L}^a(m)\,\tsf{L}_a  +  \sum_{a<b} \tsf{Q}^{ab}(m)\,\tsf{Q}_{ab}$,
\end{list}
where $\kap= N^{-1}\, \Sig^N_{n=1} \kap_n $. The second term in $\tsf{L}_{ab}(m), \tsf{Q}_{ab}(m)$ and
 $A_{ab} (m)$  subtract  the contribution resulting from the center of mass
coordinates--momenta:  $X_a= \tsf{M}^{-1}\,\Sig_n\,\tsf{m}_n\,x_{a n},\,P_\a =\Sig_{n}\,p_{a n}$.  The following formulas hold
\begin{list}{}{\itemindent=2mm\parsep=0mm\itemsep=0mm\topsep =1.5mm}
\item[$(p)$] $G[\tsf{Z}]\cdot g[\tsf{Z}]\sS=\bo 1$,  $g_{\mu \nu}[\tsf{Z}] ={\rm trace}\,\tsf{Z}_\mu\cdot \tsf{Z}_\nu$,
\item[$(r)$] $[\tsf{Z}_{\Ce}] = [\tsf{E}_{xx},\tsf{E}_{xy},\ldots,\tsf{E}_{zz}]$, $g_{(\tsf{E}_{ab},\tsf{E}_{cd})}[\tsf{Z}_\Ce]=\del_{ad}\,\del_{bc}$,
\item[$(s)$] $[\tsf{Z}_{\Re}] \sS= [\tsf{L}_x,\tsf{L}_y,\tsf{L}_z,\tsf{Q}_{xx},\tsf{Q}_{yy},\tsf{Q}_{zz},\tsf{Q}_{xy}, \tsf{Q}_{yz}, \tsf{Q}_{zx}]$,
     $g_{\mu\nu}[\tsf{Z}_\Re]\sS= ({\rm diag}\,(2,2,2,1,1,1,\tfrac 1 2,\tfrac 1 2,\tfrac 1 2))_{\mu\nu}$,
\end{list}
The most essential  points are $(d)$ and $(e)$ considered under  the assumptions $\tsf{U}_t\sS=\tsf{U}(A_t)$.
Indeed, if we put
\beqm{ \label{Ut}   (\tsf{U}_t)_{ab}= - (\pl_{A_{ba}}\,H)(\tsf{A}_t), }
then, the system at point  $(e)$ is closed and it is equivalent to the Hamiltonian equation of motion
$\dot A_\mu = \{A_\mu,H(A)\}$.
The proof of  points $(g,h)$ is elementary if the pair of bases $[Z_\Ce(m)]$ and $[\tsf{Z}_\Ce]$ is applied.
Since the coefficients of transformations $\kap$ and $\lam$ coincide, points
$(g,h)$ have to be valid for the pair ($[Z_\Re]$,$[\tsf{Z}_\Re]$), too.

Further  reduction of the set of equations of motion at the point $(f)$ and assumption (\ref{Ut}) are obtained
assuming the symmetry  of Hamiltonian $H\sS=H(A)$.

In order to discuss the reduction generated by  ${\rm SO}(3)$ invariance: $\{\tsf{L}_a,H\},\,a\sS=x,y,z$
let us introduce a mapping $\vec r \equiv (r_x,r_y,r_z) =   (x,y,z) \to [(x,y,z)] \in \tsf{K}  \subset {\rm SO}(3)$:
\begg{    [(x,y,z)]  =    \beg{pmatrix}{ 1 - a^2 & - a b  &  \bar x \\
     -\eps\,a\,b & 1 - b^2 & \bar  y  \\
    -\bar x  & - \bar y  & \bar z },     \qquad \quad \beg{matrix}{
     (a,b) = (|\vec r|\,d)^{-1/2}\,(x, y), \\ (\bar x,\bar y) = |\vec r|^{-1}\,(x,y),\\
     d  = |\vec r| +  z,  }
\intertext{and a pair of nonlinear coordinate transformations $\Lam_{\pm 1}$: }
  \label{toP1*}  \Ce^{3\times N}\setminus \Ce_{\pm } \ni  a \to  b  \in S_{\pm }\sS=
  \{  b \in \Ce_{3\times N},\,\vec{ \tsf{L}}  (\tsf{A}(\bo b,\bo b\co)) \sS=(0,0,\pm |\vec{\tsf{L}}|)\}, \\
   \label{toP*}  (b^u_{1 m},b^u_{2m},b^u_{3m} ) =
  (a^u_{x  m},a^u_{y m},a^u_{z m})\cdot [\pm \vec{\tsf{L}}(\tsf{A} (\bo a,\bo a\co))], \\
  b^1 \equiv b,\qquad b^2\equiv b\co, \\
  \Ce_{\pm }\sS=\{ a\in \Ce^{3\times N},\,\vec{\tsf{L}} (\tsf{A}(a,a\co))\neq (0,0,\mp |\vec{\tsf{L}}|)\}.  }
$\tsf{K}$ is called the Cartan model of factor space  ${\rm SO}(3)/{\rm SO}_{\vec e_3}(2)$:
\llist{  \itt{1} $[\a\,\vec x\,]\sS=[{\rm sign}\,(\a) \,\vec x]$,
         \itt{2} $[\vec x\,] \cdot \vec e_3= |\vec x|^{-1}\,\vec x$,
         \itt{3} $[\vec x\,] \cdot \vec x\cov \sS= |\vec x|\,\vec e_3$, $\vec x\cov = \trx 1 (-x_1,-x_2,x_3)$,
         \itt{4} $[\vec x\cov] = [\vec x]^{-1}$,
         \itt{5}     $[ [\vec x\,] \cdot \vec a\,] \sS= [\vec x\,]\cdot [\vec a\,]$,
         \itt{6} $[\vec x\, ] \cdot (\vec x \times \vec e_3)= \vec x \times \vec e_3$,
         \itt{7} $[\vec x\,]  = [\vec e_3\times \vec x] \cdot R_3(\a)\cdot
          [(\vec e_3 \times\vec x)\cov] \in {\rm SO}_{\vec e_3\times \vec x}(2)$,  $(\cos \a,\sin \a)\sS=
            |\vec x|^{-1}\,(\sqrt{{x_1}{^2}+{x_2}{^2}},x_3)$,      }
where ${\rm SO}_{\vec u}(2)=\{ g\in {\rm SO}(3),\,g\cdot \vec x=\vec x\}$,
$R_3(\a)= \cos\a\,(\tsf{E}_{11} +\tsf{E}_{22}) + \sin\,\a\,(\tsf{E}_{21}-\tsf{E}_{12})+ \tsf{E}_{33}$ and  $|\vec x|= (\trx 1\vec x\cdot \vec x)^{1/2}$.

Replacing $m\sS=m(a^1,a^2)\to m(b^1,b^2) = \bar m$ and ${\rm SO}(3)$ tensor
$(\tsf{L},\tsf{Q}) \to (L,Q)$  we get
\begg{  \label{Bij} B_{ij}(\bar m) =   \sum_{n}\, b^2_{in}\,b^1_{jn}= Q_{ij}(\bar m) + \tfrac i 2\,\Sig_{k}\,\eps_{ijk}\,L_k(\bar m),
       \qquad L_{i}(\bar m)\equiv 0,\,i=1,2,\\
       B_{ij} \circ \bar m(m)  =  \sum_{a b}\,[a i]\,[b j]\,A_{a b},
     \qquad [a i]= [\eps\,\vec{\tsf{ L}} (\tsf{A}(\bo a,\bo a\co))]_{m(a)i}. }
where $m(x)\sS=1,m(y)\sS=2,m(z)\sS=3$.
\ndef{   \label{u3SO}    $u^\ast_{{\rm SO}}(3)$  is algebra $\{.,.\}\toast$  spanned  by seven elements
%
% \[  (L_3,\,Q_{11}\sS=Q_1,Q_{22}=Q_2,Q_{33}=Q_3,q_1\sS=Q_{23},q_2=Q_{31},q_3\sS=Q_{12}). \]
\[  (L_3,\,Q_{1},Q_2,Q_3,q_1,q_2,q_3), \qquad Q_{i}\equiv Q_{ii},\;\; q_1\equiv  Q_{23}\,{\rm cycl}\,1, 2,3. \]
Let $B_{ij}\sS=B_{ij}\co$ and $B_{3k}=B_{k3},\,k\sS=1,2$ be the functions introduced  according to  eq.$(\ref{Bij})$.
The Poisson rules result from the following relations:
\begg{ \{B_{i_1 i_2},B_{j_1j_2} \} \toastx{\ast\,B} =
    -i\,(\del_{i_2 j_1}\,B_{i_1 j_2} - \del_{\i_1 j_2}\,B_{j_1 i_2}) \nonu  \\ +
  \label{Balg}   {L_3}{^{-1}}\,  \sum_{kl} (\gam^{kl}_{i_1 j_1}\,B_{k i_2}\,B_{l j_2}
       + \gam^{kl}_{i_2 j_1}\,B_{i_1 k}\,B_{l j_2}
       + \gam^{kl}_{i_1 j_2}\,B_{k i_2}\,B_{j_1 l}
       + \gam^{kl}_{i_2 j_2}\,B_{i_1 k}\,B_{j_1 l}),
\intertext{where ${\rm sign}\,L_3 \sS= \pm 1$  depends on the choice of a sign in the  map  $(\ref{toP1*},\ref{toP*})$  and    }
  \label{gamma}    \gam^{kl}_{ij}=  \del_{k 3}\, (\del_{j 3}\,\eps_ {i l 3} - \del_{l 3}\, \eps_{i j 3}) -
       \delta_{i 3}\, (\del_{l 3}\,\eps_{j k  3} + \del_{j 3}\, \eps_{kl3}).      }}
Formulas (\ref{Balg}) are elementary derived by the  substitution
$B_{ij}=\Sig_{k}\, b^2_{in}\,b^u_{jn}$ (see eq.(\ref{Bij}, \ref{Balg})) and assuming the following relations:
\begg{ \label{bast}   \{ b^u_{i  n}, b^v_{j m} \} \vph{|}_{b}
        = -i\,(-1)^{u-1}\,\del_{u\,3-v}\,\del_{ij}\,\del_{nm}+
          \Gam_{ij} (\vec {b}^u_{n},\vec {b}^v_{m},L_3),\\
  \label{bast1}  \Gam_{ij}(\vec a,\vec b,z) = z^{-1}\,\sum_{kl} \gam^{kl}_{ij}\,a_k\,b_l,  }
Proof of formulas (\ref{gamma},\ref{bast},\ref{bast1}) is studied in Appendix.
Algebra $u^\ast_{{\rm SO}}(3)$ decomposes $u^\ast_{{\rm SO}}(3)= {\rm su}^\ast_{{\rm SO}}(3) \times C_1$
 where
$C_1={\rm trace}\,\tsf{B} \sS=\Sig_{i\le 3 }\,Q_i\sS=Q_1+Q_2+ Q_3$ is the centrum
while $ {\rm su}^\ast_{{\rm SO}}(3)\sS=\{L,\bar Q_1,\bar Q_2,q_1,q_2,q_3\}$
where $\bar Q_1=2^{-1/2}\,(Q_1-Q_2),\bar Q_2=6^{-1/2}\,(2\,Q_3-Q_1-Q_2)$.
For the matrix $\tsf{B} =\Sig_{ij}\, B_{ij}\,\tsf{E}_{ij}$ we get
\begg{ \tsf{B} =  \beg{pmatrix}{ Q_1 & q_3+ \tfrac {i}2 L_3 & q_2 \\
        q_3- \tfrac {i}2 L_3 & Q_2 &  q_1 \\ q_2 &q_1 & Q_3  }, }
%
%  \beg{matrix}{  Q_1= p_+ + Q_+ + Q_- , \\    Q_2= p_+ + Q_+ - Q_- ,\\
 %  Q_+ = \tfrac 1 2\,(\Sig_{i\le 3}\, p_i - Q_3),\qquad   Q_i = Q_+  +(-1)^{i-1}\,Q_-, \qquad i\sS=1,2,}
%
where $q_{3}\equiv Q_{12},\,{\rm cycl}\,1,2,3$.
From formulas (\ref{Bij},\ref{Balg})  we get
%
% \begin{comment}
\Meq{su3crul}{
\begg{ %\label{su3crul}
  \{Q_k,L_3\}\toast = 0, \qquad \{q_3,L_3\}\toast=Q_1- Q_2,\qquad
      \{q_i,L_3 \}\toast = (-1)^{i-1}\,q_{3-i}, \\
     \{Q_i,Q_j\}\toast = -4\,(\Sig_k \eps_{ijk})\,L^{-1}\,q_1\,q_2, \\
     \{q_1,q_2\}\toast =  {L_3}{^{-1}}\,[ (Q_1-Q_3)\,(Q_2-Q_3) - {q_3}^2 - \tsf{G}\, {L_3}{^2}/4], \\
     \{q_i,q_3\}\toast = {L_3} ^{-1}\,(-1)^{i-1}\,[q_{3-i}\,q_3 + q_i\, (Q_3-Q_{3-i}) ],   \\
%
%    \{q_i, Q_3 \}\toast = \frac {(-1)^{i}\,2}{L_3 }\,[ q_i\,q_3+ ( Q_3-Q_{3-i})\,q_{3-i} ], \\
%
        \{Q_i,q_j\}\toast= \frac{2}{L_3}\,\beg{pmatrix}{
        q_2\,(Q_2-Q_3) & q_2\,q_3 & \tfrac {\tsf{G}\,{L_3}^2}4 -{q_2}^2 \\
        q_3\,q_1 & q_1\,(Q_3-Q_1) & {q_1}{^2} - \tfrac {\tsf{G}\,{L_3}^2}4\\
        q_3\,q_1+(Q_3-Q_2)\,q_2 & - q_2\,q_3 -(Q_3-Q_1)\,q_1 & -{q_1}{^2}-{q_2}{^2} }_{ij}. }}
%
%  if $\tsf{G}\sS=1$ then  point $(h)$ is fulfilled,  hence $
%
% \end{comment}
where the   Poisson rules for  coalgebra $u^\ast_{\rm SO}(3)$ we find   putting  $\tsf{G}\sS=1$.
If $L_1\equiv L_2\equiv 0$ then
$\{L_1,x\}\toast=\{L_2,x\}\toast=0$ are fulfilled for all $x\in u^\ast_{\rm SO}(3)$ identically.
 %  $\tsf Q\equiv \tsf Q_{\tsf{G}=1}$.More precisely $ |_{\tsf{G}=1}, \tsf Q\sS= \tsf Q(Q,\tsf L)$

The case $\tsf{G} \sS=0$   is also  physically interesting. Assuming $Q|_{\tsf{G}\sS=0} =\c Q$
one finds  $\{\c Q_{ab},\c Q_{cd}\}\sS=0$ and
$\{\tsf L_a,\c Q_{bc}\}= \sum_{d}\,(\eps_{abd}\,\c Q_{dc}+\eps_{acd}\,\c Q_{bd})$; hence, $(\tsf  L,\c Q)$ is the semidirect
Poisson algebra  obtained  considering the mass quadrupole--monopole tensor:
 $\c Q= \Sig_{k} \tsf{m}_k\,\vec x_k \otimes  \vec x_k - \tsf{M}\,\Vec X \otimes \Vec X$
where  the   center of mass  reference   frame  is applied.
\section{ \label{Orbits} Canonical coordinates on $[p_1,p_2,p_3]_\ast$ orbits.}
Coadjoint  orbits of $u\co(3)$ coalgebra are found studying  a surface  ${\rm U}(3) \ni g  \to
\tsf{A}(g\bo \cdot  m_0)= {\rm Ad}_g (m_0)$.
These orbits denoted as   $[p_1,p_2,p_3]$ are labeled by components of the weight vector $\vec p=(p_1,p_2,p_3)$
accordingly with  the following Casimir relations:
\begg{ \label{Cas} C_k(\tsf{A})\sS= S_k(\vec p),  \qquad
  C_{k}(\tsf{A})\sS= {\rm trace}\, \tsf{A}_k, \qquad \tsf{A}_1\sS=\tsf{A},\quad \tsf{A}_{k+1}= \tsf{A}_k\cdot \tsf{A}_1, }
where $S_k(\vec p)\equiv S_k$.  Here $S_k$, as well, a few other ones
\begg{   \label{Sk} S_k \equiv  \Sig_{i\le 3}\, {p_i}{^k}, \qquad \\
      S_{k1}= (1+\del_{k1})^{-1}\,\Sig_l (p_{l+1}+p_{l+2})\,{p_l}{^k}, \\
      S_{111}=p_1\,p_2\,p_3, \quad  S^B_{111}= (p_1-p_2)\,(p_2-p_3)\,(p_1-p_3),\\
      S^A_{111}= (p_1+p_2- 2\,p_3)\,(p_2+p_3-2\,p_1)\,(p_1+p_3-2\,p_2).    }
 are frequently used   functions of $\vec p$. The
A peculiar  class of orbits will be discussed here. They are obtained using  the assumptions:
$p_1\sS> p_2 \sS> p_3$.

The ${\rm SO}(3)$ reduction of $u\co(3)$: $u\co(3)\to u^\ast_{\rm SO}(3)$ constraint  $[p_1,p_2,p_3]$,
${\rm dim} [p_1,p_2,p_3]= 6$ to four dimensional orbits $[p_1,p_2,p_3]\toastx{\ast \hat \eps}$. %
Casimir functions $C_k$  for  $k\sS=1,2,3$ are independent.

Choosing $(q_1,q_2,Q_3,L_3)$ as a set of independent coordinates,
let us rewrite three Casimir relations (see, eq.(\ref{Cas},\ref{Sk})) in the following form:
\begg{  \label{CASR}  Q_+ \sS= \tfrac 1 2\,(S_1  - Q), \qquad
       \beg{pmatrix}{ {q_1}^2 -{q_2}^2  & 2\,q_1\,q_2 \\ -2\,q_1\,q_2  & {q_1}^2 -{q_2}^2 }
        \binom{q_3 }{Q_-} = \binom{V_{L}(Q,R) }{U(Q,R) }.
\intertext{where }
    U(Q,R) \sS =       G_{\vec{p}}(Q) + \tfrac 3 2\,(Q-\la p\ra)\,R^2, \\
    V_{L}(Q,R) = \eps\,[-\hat w_{\vec{p}}(L,Q,R)]^{1/2}.}
and $\eps\sS=\pm 1$.
The following formulas have been applied
\begg{  Q \equiv Q_3, \qquad Q_\pm \equiv \tfrac 1 2\,(Q_1\pm Q_2), \qquad L\equiv L_3, \qquad
  R^2\equiv {q_1}^2 +{q_2}^2,  \\
  \la p\ra =\tfrac 1 3\,S_1,\qquad G_{\vec{p}}(x) = \prod_{i} (x-p_i),\\
  \hat w_{\vec{p}}(a,b,c)= \tfrac 1 4\,b^4\,a^2 + H_{\vec{p}}(b,c), \\
   H_{\vec{p}} (b,c)= \prod_{i<j} h_{ij}(b,c),\qquad h_{ij} (b,c) = (b-p_i)\,(b-p_j)+c^2, }
New coordinates $(p,\gam)$ are determined introducing a pair of relations
\beqm{ \label{pGmix}
       {R_{L,Q,p}}{^2} = - \frac {G_{\vec{p}}(Q)} {Q-p},\qquad
       \cos^2 \gam_{L,Q,p}= \frac {L^2}{4} \,\frac{Q-p}{G_{\vec{p}}(p)}.    }
They define mapping
 $\Lam_L \!: (p,\gam)\to (Q,R)=(Q_{\vec p,L}(p,\gam),R_{\vec p,L}(p,\gam))$ where
\begg{ \label{RL}   R_{\vec p,L}(p,\gam) = (2\,L^2\,|\cos \gam|)^{-1}\,\sqrt{-F_{\vec p,L}(p,\gam)}, \\
      \label{QL}      Q_{\vec p,L}(p,\gam)\sS= p +4\,L^{-2}\,G_{\vec{p}}(p)\,\cos^2 \gam,\\
     \label{toF}    F_{\vec p,L}(p,\gam) =
   \prod_{i\le j}\,\hat h_{ij}(L,p,\gam), \qquad \hat h_{ij}(a,b,\gam)=a^2 + 4\,(p_i-b)\,(p_j-b)\,\cos^2 \gam.
\intertext{Simple calculations lead to the following two identities}
              H_{\vec{p}}\circ \Lam(p,\gam)= -(2\,L\,\cos \gam)^{-6}\,{F_{\vec{p}}}{^2} (L,p,\gam)=
            -\frac {L^2\,{R_{\vec p,L}}{^4}(p,\gam)}{4\,\cos^2 \gam}, \\
      \label{VL}   V_{\vec p,L}(p,\gam)=  V_L \circ \Lam_L(p,\gam)=
     \tfrac{1}8\,L^{-3}\,F_{\vec p,L}(p,\gam)\,\cos^{-3}(\gam)\,\sin \,\gam,}
where  the function  $F_{\vec p,L}(Q,R)$ (presented in eq.(\ref{toF})):
$(a)$ is negative valued on $\tsf{M}^\pm_{\vec{p}}$ (see, eq.(\ref{RL})) and
$(b)$ is  an  even function of $\gamma$; so $V_{\vec p,L}(p,\gam)$ is
\begin{list}{}{\itemindent=2mm\parsep=0mm\itemsep=0mm\topsep =0mm}
\item[$(a)$]   an odd function of  $\gam$,
\item[$(b)$]  ${\rm SO}(3)$ scalar function,
\end{list}
and,
\begin{list}{}{\itemindent=2mm\parsep=0mm\itemsep=0mm\topsep =0mm}
\item[$(c)$]  ${\rm sign}\,\sin\,\gam \sS= - {\rm sign}\,V_L=- \eps$.
\end{list}
\par
Employing point $(c)$ we get the following form of the inverse transformation
$(Q,R) \to (p,\gam)\sS=(p_{\vec{p}}(Q,R),\gam_{L,Q,R})$:
\begg{   p_{\vec{p}}(Q,R) = Q+ R^{-2}\,G_{\vec{p}}(Q),\\
   \label{toGamQ}
    \binom{\cos \gam_{L,Q,R}}{ \sin \gam_{L,Q,R}  }  = \frac{1} {\sqrt{-H_{\vec{p}}(Q,R)}}\times
    \binom{ \frac{L\,R^2}{2}}{ - V_{L}(Q,R)  }. }
where $\eps_\gam \sS= {\rm sign}\,\gam$, and we assumed $\sqrt{\smash[b]{|x|}}\ge 0$.
%
% With help of $\Lam_L$
%
Let us define  $\Gam  \!: \tsf{M}^{\hat \eps}_{\vec{p}} \ni (L,p,\vp,\gam) = m  \to
\Gam (m)= (q_1,q_2,q_2,Q_1,Q_2,Q_3, L_3)\in [ \vec p ] \toastx{\ast \hat \eps}$ where
\Meq{GAM}{
 \begg{   L_3 \sS= L, \qquad Q_1\sS=Q_+ +  Q_-, \quad Q_2= Q_+ - Q_-,\qquad  Q_3\sS= Q_{\vec p,L}(p,\gam), \\
     Q_+ = \tfrac 1 2\,[S_1  - Q_{\vec p,L}(p,\gam)], \\
      q_{1} +  i\,q_2  = e^{-i\,\vp}\,(2\,L^2\,\cos \gam)^{-1}\,\sqrt{ \smash[b]{-F_{\vec p,L}(p,\gam)}}, \\
      Q_- + i \, q_3 =  e^{ i\,2\,\vp}\,[\tfrac 3 2\,(p-\la p\ra) - \tfrac i 2\,L\,\tan \gam+
         2\,L^{-2}\,G_{\vec{p}}(p)\,\cos^2\gam], }}
The sets  $\tsf{M}^{\hat \eps} _{\vec{p}} = \Gam\ord{-1} ([\vec p] \toastx{\ast \hat \eps})$ are discussed below.
Signatures $\hat \eps ={\rm sign}\,L_3$ result from the choice of
${\rm SO}(3)$ matrices $[\hat \eps\,\vec L]$  determining the coordinates $b$ (see, eq.({\ref{toP*})).
\ntheo{ \label{Omth} $\Gam (\tsf{M}^{\hat \eps}_{\vec{p}}) = [\vec p] \toastx{\ast {\hat \eps}}$ for
 $\tsf{M}^{\hat \eps}_{\vec{p}}\sS=  \bigcup_{0<\,\hat \eps\,L}
    \tsf{M}_{2,\vec p,L} \setminus (\overline{\tsf{M}_{3,\vec p,L}} \cup \overline{ \tsf{M}_{1,\vec p,L}}) $
where
\begg{ \tsf{M}_{i,\vec p,L}=   \{ (L,p,\vp,\gam),\; \vp \equiv \vp\,{\rm Mod}(2\,\pi),\,0< \cos \gam,\,
       p^-_{i,\vec p}\,(L\,\sec \gam) < p <  p^{+}_{i,\vec p}\, (L\,\sec \gam)\}, \nonu \\
      p^\pm_{i,\vec p}(l)\sS=  \frac 1 2\,\left(  p_{i+1}+p_{i+2} \pm \sqrt{(p_{i+1}-
      p_{i+2})^2-l^2}\right), \quad  p_{k} \equiv p_{1+ {\rm Mod}_3(k)}, \nonu }
The sets $\overline{\tsf{M}^{\hat \eps}_{i,\vec p}}$ closes  $\tsf{M}^{\hat \eps}_{i,\vec p}$ where
$\overline{\tsf{M}^{\hat \eps}_{i,\vec p}}= \bigcup_{\hat \eps\,L>0}\overline{\tsf{M}^{\hat \eps}_{i,L,\vec p}},\,
  \overline{ \tsf{M}^{\hat \eps}_{i,L,\vec p}} \sS= \{(L,p,\vp,\gam),\,
   \vp \equiv \vp\,{\rm Mod}(2\,\pi),\,0< \cos\,\gam,\;
   f_i(L,p,\gam) \le  0\}$ and $f_1(L,p,\gam)=\hat h_{23}(L,p,\gam),\,{\rm cycl}\,1,2,3$ $[$see eq.$(\ref{toF})]$.
\\ \hph{} \hskip \parindent
The pairs $(\vp,L)$ and $(\gam,p)$ are mutually commuting canonical coordinates such that
\beqm{  \Om_{M^\pm_{\vec{p}}}(m)\sS= d L\we d\vp \mp   dp \we d\gam,  }}
\par
If $l\le {\rm min}\,(p_1-p_2,p_2-p_3)$ then  $p^\pm_{i,\vec p}(l)\in \Re$ and
 $p^-_{2,\vec p}(l) \sS\le p^-_{1,\vec p}(l) < p^+_{1,\vec p}(l) \le p^-_{3,\vec p}(l)<
 p^+_{3,\vec p}(l) \sS \le p^+_{2,\vec p} (l)$. \par
The reduction of range for $\gam\!:\,\cos\,\gam\sS>0$ results
from the identity:  $\Gam(L,p,\vp+\pi,\gam+\pi)= \Gam(L,p,\vp,\gam)$. \par
Subspaces $\tsf{M}^\sig_{i,\vec p,}$ obey the following rules:
\begin{list}{}{\itemindent=2mm\parsep=0mm\itemsep=0mm\topsep =0mm}
\item[$(a)$]  $\tsf{M}^{\hat \eps} _{2,\vec p,L} \neq \emptyset \Leftrightarrow \hat \eps\,L\le p_1-p_3$,
\item[$(b)$]  $\tsf{M}^{\hat \eps}_{3,\vec p,L}\neq \emptyset \Leftrightarrow \hat \eps\,L\le p_1-p_2$,
\item[$(c)$]  $\tsf{M}^{\hat \eps}_{1,\vec p,L}\neq \emptyset \Leftrightarrow \hat \eps\,L\le p_2-p_3$,
\item[$(d)$] $\tsf{M}^{\hat \eps} _{2,\vec p,L}\cap \tsf{M}_{i,\vec p,L}\sS=\tsf{M}_{i,\vec p,L}$
\item[$(e)$] $\tsf{M}^+_{i,\vec p,L}\cap \tsf{M}^-_{j,\vec p,L} = \emptyset$.
% \item[$(g)$] $( R_{\vec p,L}(\pl \tsf{M}_{i,\vec p,L}),Q_{\vec p,L}(\pl \tsf{M}_{i,\vec p,L}))= (0,p_i)$.
\end{list}
{\em Proof of theorem} \ref{Omth}. Applying to formulas $(\ref{su3crul})$  the rules of coordinate transformation
obtained from mapping  $(\vp,R) \to (q_1,q_2)$ and choosing the coordinates $(q_1,\ldots,q_4)=(\vp,L,Q,\\ R)$
one finds
\begg{  \{\vp,L\}\toast=1,\qquad
  \{\vp,Q\}\toast = 2\, L^{-1}\,R^{-2}\,G_{\vec{p}}(Q), \nonu \\
  \{\vp,R\}\toast =  (L\,R)^{-1}\,[3\,Q^2 +  R^2 - 2\,S_1\,Q + S_{11} + \tfrac 1 4\,(1-G)\,L^2], \nonu \\
   \{L,Q\}\toast=\{L,R\}\toast =0, \qquad   \{Q,R\}\toast= 2\,L^{-1}\,R^{-1}\,V_{L}(Q,R). \nonu }
%
\begin{comment}
\Meq{su3crul}{
\begg{ %\label{su3crul}
  \{Q_k,L_3\}\toast = 0, \qquad \{q_3,L_3\}\toast=Q_1- Q_2,\qquad
      \{q_i,L_3 \}\toast = (-1)^{i-1}\,q_{3-i}, \\
%
     \{Q_i,Q_j\}\toast = -4\,(\Sig_k \eps_{ijk})\,L^{-1}\,q_1\,q_2, \\
     \{q_1,q_2\}\toast =  {L_3}{^{-1}}\,[ (Q_1-Q_3)\,(Q_2-Q_3) - {q_3}^2 - \tsf{G}\, {L_3}{^2}/4], \\
     \{q_i,q_3\}\toast = {L_3} ^{-1}\,(-1)^{i-1}\,[q_{3-i}\,q_3+ (Q_3-Q_{3-i})],   \\
%
%    \{q_i, Q_3 \}\toast = \frac {(-1)^{i}\,2}{L_3 }\,[ q_i\,q_3+ ( Q_3-Q_{3-i})\,q_{3-i} ], \\
%
        \{Q_i,q_j\}\toast= \frac{2}{L_3}\,\beg{pmatrix}{
        q_2\,(Q_2-Q_3) & q_2\,q_3 & \tfrac {\textsf{G}\,{L_3}^2}4 -{q_2}^2 \\
        q_3\,q_1 & q_1\,(Q_3-Q_1) & {q_1}{^2} - \tfrac {\textsf{G}\,{L_3}^2}4\\
        q_3\,q_1+(Q_3-Q_2)\,q_2 & - q_2\,q_3 -(Q_3-Q_1)\,q_1 & -{q_1}{^2}-{q_2}{^2} }_{ij}. }}
%
\end{comment}
Let $ \hat \om^{-1}= \om $ where
$\hat \om\!:\,\hat \om_{ij}\sS=\{q_i,q_j\}$.  Then
$\Om_G(\vp,L,Q,R)= \sum_{i\le j}\,\om_{ij}\,dq_i\we dq_j$ is a symplectic two--form.
The explicit calculation gives
\begg{  \Om_{\textsf{G}} (\vp,L,Q,R)=   dL\we (d\vp + \om_{LQ}\,dQ+ \om_{LR}\,dR)+\om_{RQ}\,dR\we dQ, \\
       \om_{LR}= (V_L\,R)^{-1}\, G_{\vec{p}}(Q), \\ \om_{RQ} = \tfrac 1 2\;{V_{L}}{^{-1}}(Q,R)\,L\,R, \\
      \om_{LQ,\textsf{G}}=  (2\,V_{L})^{-1}(Q,R)\,[ -  3\,Q^2 + 2\,S_{1}\,Q -  S_{11}- R^2 + \tfrac 1 4\,(\textsf{G}-1)\,L^2]. }
With the help of relations
 \begin{list}{}{\itemindent=2mm\parsep=0mm\itemsep=0mm\topsep =0mm}
\item[$(a)$]  $dp= (1+ R^{-2}\, G\pr_{\vec{p}}(Q))\,dQ - 2\,R^{-3}\,G_{\vec{p}}\,dR$,
\item[$(b)$]  $d\gam= -\cot\,\gam\,[   L^{-1}\,dL   - \tfrac 1 2\,\,{H_{\vec{p}}}{^{-1}}\,H\ord{1,0}_{\vec{p}}\,dQ + (2\,R^{-1}-
   \tfrac 1 2\,{H_{\vec{p}}}{^{-1}}\,H\ord{0,1})\,dR ]$,
\item[$(c)$] $\cot \gam = -\tfrac 1 2\,  {V_{L}}{^{-1}}\,L\,R^2$,
\item[$(d)$] $dp \we d\gam =  {V_{L}}^{-1}\,[\,R^{-1}\,G_{\vec{p}}\;dR \we dL  +
\tfrac 1 2\,R^2\,(1+R^{-2}\,G\pr)\,dL\we dQ +
      u\,dQ \we dR]$,
\item[$(e)$] $u \sS=  ( 4\,H_{\vec{p}}\,R)^{-1}\,L\, [2\,G\,H \ord{1,0}_{\vec{p}} -
  4\,H\,(R^2+(G_{\vec{p}})\pr)+ R\,H \ord{0,1}_{\vec{p}}(R^2+(G_{\vec{p}})\pr)] \sS=       \tfrac 1 2\,L\,R,$
\end{list}
we find $dp\we d\gam= \om_{LR}\,dL\we dR + \om_{LQ,1}\,dL\we dQ + \om_{RQ}\,dR\we dL$ which proves the statement.
\section{Ellipsoids}
The ${\rm su}^\ast_{\rm SO}(3)$ Hamiltonians  are ${\rm SO}(3)$ invariant if $0\sS\equiv H_{,\vp}\Rightarrow
\bar H\sS=H_{L}(p,\gam)$. The function $V_L$ commutes
with angular momentum: $\{V_L,\,L\}\toast\sS=0$; hence, the most general form of the function $H_L(p,\gam)$ is
obtained  using (in general independent) two functions $\bar h_{i,L}(Q,R)$
\begg{   H_L(p,\gam)=  h_{1,L}(p,\gam ) +  h_{2,L}(p,\gam)\times V_{L}\circ \Gam(p,\gam),\\
        h_{i,L}(p,\gam)= \bar h_{i,L}\circ \Gam_L(p,\gam), }
where the second term of $H_L(p,\gam)$ represents    $\gam$ odd contribution.
Since
\beqm{  R_{,L}(p,\gam) \equiv
  (\pl_L R_{\vec p,L})(p,\gam) = L^{-1}\,{R_{\vec p,L}}^{-1}(p,\gam)\times(R^2+3\,Q^2 - 2\,S_1\,Q+S_{11}), }
$(\pl_L\, h_{i,L})(Q,0)$ is finite only  if $|\lim_{R\to 0} R^{-1}\,h\ord{(0,1)}_i(Q,R)| \sS< \infty $.
\par
Let
\beqm{ \Om_{\vec p,L}(p,\gam)\sS=H_{L,L}(p,\gam), \qquad \Lam_{\vec p,L}(p,\gam)= H_{L,p}(p,\gam), }
denote first derivatives of Hamiltonian $H_{L}(p,\gam)$.
\ndef{ Points $m=(L,p,\vp,\gam_{i,p})$  of sets $\pl \tsf{M}_{i,\vec p,L}$
are selected from the following conditions:
\begg{   p\in [p^-_{i,\vec p}(L),p^+_{i,\vec p}(L)], \qquad  L\in [0,p_{i>}-p_{i<}], \qquad \binom{p_{i<}}{ p_{i>} } =
 \binom{ {\rm min}(p_{i+1},p_{i+2})}{ {\rm max}(p_{i+1},p_{i+2})   }, \\
    \cos \gam_{i,p} \sS=\frac{L}{2\,\sqrt{p_{i>} -p}\,\sqrt{p-p_{i<}} } ,  \qquad i=1,2,3,} }
We have:
\begin{list}{}{\itemindent=2mm\parsep=0mm\itemsep=0mm\topsep =0mm}
\item[$(a)$] $Q\sS=Q(m) = Q_L(p,\cos \gam_{i,p})=  p_i$.
\item[$(b)$] $R_{\vec p,L}(p,\gam_{i,p})\sS=0$.
\end{list}
\ndef{ The states $\pl \tsf{M}_{i,\vec p,L}$ are called  $\tsf{S}_{\tsf{i}}$--ellipsoids.
The condition $\Lam_{\vec p,L} (p,\gam) \sS= 0$  selects the family of $\tsf{P}$ ellipsoids.}
Here and further  physical  states  $m$ will be described using the map
  $\tsf{M}^+_{\vec{p}}, (\tsf{M}_{i,\vec p} \equiv \tsf{M}^+_{i,\vec p})$:  $L\sS=L(m)>0$.

$\tsf{S}_{\tsf{i}}$ ellipsoids:
$L\sS=p_{i>}  -  p_{i<}\sS= L_{i,\rm max},\,i=1,2,3$  we call the maximal states.
For  maximal states
\begin{list}{}{\itemindent=2mm\parsep=0mm\itemsep=0mm\topsep =0mm}
\item[$(a)$] $p^-_{i,\vec p}(L_{i,\rm max})\sS= p^+_{i,\vec p}(L_{i,\rm max})= \tfrac 1 2\,(p_{i<} +p_{i>})$
\item[$(b)$] $\gam \sS=\gam_p \sS = 0$,
\item[$(c)$]  $q_1-q_2\sS=Q_-\sS=0 \Rightarrow \Gam(L_{i,\rm max},\tfrac 1 2\,(p_{i<} +p_{i>}),\vp+\a,0) =
  \Gam(L_{i,\rm max},\tfrac 1 2\,(p_{i<} +p_{i>}),\vp,0)$.
\end{list}
i.e., they are {\em axially--symmetric states} (see  point $(c)$).
\par
The discussion of the family of $\tsf{P}$ ellipsoids  becomes  much simpler in the cases when
$H(L,p,\gam)$ is an even function of coordinates $\gam$:  $H(L,p,\gam)\sS=H(L,p,-\gam)$. Since
\beqm{ \label{Pballance}    H_{,p}(L,p,0) =0, \qquad p\neq p_i,\;i=1,2,3,   }
these extremes exist   for $\gam \sS=0$.
As a natural example let us discuss $\tsf{P}$--ellipsoids for the  following  Hamiltonians:
\beqm{ \label{HAM}  H_{\om,g}(L,Q) \sS= E_0 + \tfrac 1 4\,\om(L)\,Q, }
The physical role  of this family is exhibited by the following formula:
\begg{  {\rm det}\,\tsf{Q} = p_1\,p_2\,p_3 + \tfrac 1 4\,Q\,L^2.
\intertext{The simple subfamily is derived   considering the functions}
   \label{HAMM}     H_{\om,r,s}(\tsf{Q} )  =
     \kap^{-1}(s)\,\om_{\la r \ra}\,\geo{p}{1- 3\,\kap(s)}\,{\rm det}^{\kap (s)}\,\tsf{Q}, \\
  \qquad   \kap(s) = \frac {6\,s}{s+2},  \qquad   \om_{\la r\ra} =  (p_V/\geo{p}{} )^{r-1}\times \om,  }
where $\om$ is the nuclear constant $\hbar\, \om =\hbar\,\om_\tsf{A} \approx 40\,\tsf{A}^{-1/3}\,{\rm MeV}$,
$\geo{p}{}\sS=  (p_1\,p_2\,p_3) ^{1/3} \approx p_{V}$ where $p_V \equiv \tfrac 1 6\,(3\,\tsf{A}/2)^{4/3}\,\hbar$
is simple estimation of the Pauli selection rule for a neutron--proton system
resulting from application of  the triaxially  deformed  harmonic oscillator  shell model.
Assuming $\om(L)\ll  \om_V$, then preserving the linear term and comparing the result with formula, eq.(\ref{HAM}),
we get
\begg{  \om(L)  \approx  \geo{p}{-2}\,\om_{\la r \ra}\,L^2,
\intertext{as well}
  \label{Omi}  \Om_{i,r}(0) = (\pl_{p_i}  H_{\om,s,r})(0)   =
  \frac {(2-r)\,(2+s)}{2\,s} \times \frac {\om_{\la r \ra}\,\geo{ p}{}}{p_i},}
where $\geo{x}{} \sS= (x_1\,x_2\,x_3)^{1/3}$.
As in the physical  model $\Om_{i,r}(0)$ has to be positive,
the physical  range of parameters $(s,r)$ is limited by the conditions: $r\le 2$ and $0 \le s$.

If  $r\sS=r_{\rm h.o.}$, where $r_{\rm h.o.}\sS=4/(2+s)$, then
$\Del E_{\vec{p}}(0) \sS=E_{\vec p+\del \vec p}=
 \Sig_i \,
 \Om_{i,r}(0)\,\del p_i + \ldots$ is power series of the excitation energy.
%
\begin{comment}
Quantum--mechanically, $\del p_i$ express the modification of the single particle occupation quantum numbers for
particles  moving in the selfconsistent harmonic oscillator field:
$U(L) \sS= m\,\Sig_{in}\,{\Om_{i,  r_{\rm h.o.}  }}{^2}(L) \times {x_{in}}{^2}$.
\end{comment}
One finds
 \begg{   \Om_{\vec p,L}(p,\gam) = \frac {s\,p\,L\,H_{\om,r,s}}{3\,(2+s)} , \quad
    \Lam_{\vec p,L}(p,\gam) = \frac {s\,[L^2 - {\c L_{\vec{p}}}{^2} (p)\,\cos^2\gam ]}{6\,(2+s)}\,H_{\om,r,s},\\
            \c L_{\vec{p}}(p)= 2\,\sqrt{- 3\, p^2 + 2\,p\,S_1 - S_{11}}. }
The functions $\c L_{\vec{p}}(p)$ fulfill the following rules:
\begin{list}{}{\itemindent=2mm\parsep=0mm\itemsep=0mm\topsep =0mm}
\item[$(a)$]  $ \c L_{\vec{p}}\,(p_\pm(L)) \equiv L$,
\item[$(b )$]  $p_\sig(\c L_{\vec{p}}(p))= p $ if $\sig\,(S_1-3\,p)>0$,
\item[$(c)$]  $p_\sig(\c L_{\vec{p}}(p))= \tfrac 1 3\,( 2\,S_1- p)$  if $\sig\,(S_1-3\,p)<0$,
\end{list}
where  $\sig\sS=\pm 1$ and
\begg{  \label{pellP}   p_\pm (L) =  \tfrac 1 3\,\bigl( S_1 \mp  S_L), \qquad
     S_L= \tfrac 1 2\,\sqrt{-3\,L^2+4\,(S_2-S_{11})}.}
The conditions $p = p_{\pm}(L,\vec p)$  select  the family of $\tsf{P}_+$  and $\tsf{P}_-$ ellipsoids, respectively.

The physical interpretation of the functions $p_{\pm}(L)$ is provided by the following theorem:
\ntheo{ \label{Qtheo}  Tensor $\tsf{Q}$ possesses degenerated eigenvalues
 $P_1\sS=P_2\ge P_3$ if and only if $p=p_+(L)$ and $P_1> P_2\sS=P_3$ if and only if
$p=p_-(L)$.    }
In order to  prove theorem \ref{Qtheo},  we have to  check the validity of the following rules:
\begin{list}{}{\itemindent=2mm\parsep=0mm\itemsep=0mm\topsep =0mm}
\item[$(a)$] formula (\ref{Pi}),
\item[$(b)$] the relation  $V_{\vec{p}}(L,Q,x)\sS=-\tfrac 1 4\,L^2 (Q_{L,\vec p}(x,0)-Q)$  (see also, eq.(\ref{V})), and,
\item[ $(c)$] pair of identities: $(\pl_x Q_{L,\vec p})(x,0)|_{x=p_{\pm}(L)}=0 $.
\end{list}
The explicit expressions for  eigenvalues $P_\a$ found
from equation (\ref{Pi}) are studied in a number of relations (\ref{P123a}--\ref{P123d}).
Let $\tsf{D}_{\vec{p}}$  denote the set obtained from projection of $\tsf{M}_{\vec{p}}$ onto the plane $(L,Q)$.
Using formulas, eq.(\ref{P123d}),  we find
\begg{  \tsf{ D} _{\vec{p}} =   \bigcup_{0\le L\le p_1-p_3} (L,[Q^L_{\rm min},Q^L_{\rm min}]), \\
     \label{QintL}     Q^L_{\rm min}=
  \beg{cases}{ p_3 &  0 \le L\le \lam, \\
              \bar Q_{-}(L) &  \lam \le L \le h(\lam,\mu), \\ p_2 & h(\lam,\mu) \le L\le \lam+\mu,  }  \qquad
         Q^L_{\rm max} = \beg{cases}{ p_1  &  0 \le L\le \mu, \\
         \bar Q_{+}(L) & \mu \le L \le h(\mu,\lam), \\
         p_2 & h(\mu,\lam)\le L \le \lam+\mu }      } %
\begg{  \label{QellP}        \bar Q_\sig (L)\sS=  Q_{L,\vec p}(p_\sig(L),0),\\   h(x,y)= \sqrt{(x+y)^2 - {\min}^2(x- y,0)}=
         \beg{cases}{ x +y &  x \ge  y, \\ 2\,\sqrt{x\,y} & {\rm else.} }  }
where $\bar Q_\sig(L)$ obey the following identity:
$ V_{\vec{p}}(L,\,\bar Q_\sig (L), p_\sig(L)  )\equiv 0$ for  $V_{\vec{p}}(L,Q,p)$ given  in eq.(\ref{V}).
Formulas, eqs.(\ref{pellP},\ref{QellP}),  separate on
orbits $[ \vec p]\toast$  two bands $\tsf{P}_-$ and $\tsf{P}_+$:
\begg{ \label{topL}  \tsf{P}_{\pm}\!: \quad \gam=0, \qquad    p\sS=p_{\pm }(L),\quad Q=\bar Q_\sig(L).  }
Angular momentum   range  in the case of  $\tsf{P}_-$ band   is equal
to $L\in [\lam,h(\lam,\mu)]$, while  for $\tsf{P}_+$ band:  $L\in [\mu,h(\mu,\lam)]$ (see, eq.(\ref{QintL}).

The function $S_L$ is real if  $L^2 \le  \bar L^2\sS=
\frac 4 3\,(S_2-S_{11}) \sS={L_{\rm max}}^{2} + \tfrac 1 3\,(\lam-\mu)^2$ where $L_{\rm max}\sS=\lam+\mu$
is maximal physical  value of angular momentum. Hence, $S_L$ is positive valued function.
One finds
\begg{  \bar E_i =   {\rm det}\,\tsf{Q}(\tsf{S}_{\tsf{i}}) =   p_i\,(p_{i+1}\,p_{i+2}+ \tfrac 1 4\,L^2), \\
   \bar E_{\pm} =  {\rm det}\,\tsf{Q}(\tsf{P}_\pm) = S_{111} +
  \tfrac 1 {27}\,S^A_{111} + \tfrac {1}{4}\,\la p\ra\,L^2  - \tfrac 1 {18}\,(p_\pm(L,\vec p)-\la p\ra)^{3/2},  }
% \beqm{ \pm \tfrac 1 {108}\,[4\,(S_2-S_{11}) - 3\, L^2]^{3/2},
%
where by $\bar E_3 <\bar E_2 <\bar E_1$ we denoted values of  the energy factor
${\rm det}\,\tsf{Q}$  in  the cases  of   $\tsf{S}_{\tsf{i}}$ ellipsoids, while
$\bar E_{-}<\bar E_{+}$ determine the  energy factor in the case of  $\tsf{P}_-$  and $\tsf{P}_+$ ellipsoids, respectively.

It is interesting to compare the derived  expression onto solutions for  $\tsf{S}_{\tsf{i}}$ bands
with the earlier study\cite{cerkAn} of the model of $\tsf{S}$--ellipsoids
based on the orbits $\la p_1,p_2,p_3\ra$  of ${\rm sp}\co(6,R)$ Poisson algebra.
For the many--particle system bounded by the ${\rm SO}(N)\times {\rm SO}(3)$ invariant potential:
$U(\c Q)\sS= (s+2)^{-1} \,H_{\om,2,s}(m\,\om_\tsf{A}\,\c Q)$ the system of condition selecting
$\tsf{S}_{\tsf{i}}$ ellipsoids reduces.
In the limit $p_{i+1}-p_{i+2}\ll p_{i+1}+p_{i+2}$ the asymptotic formulas read%
\beqm{ \label{dQ} {\rm det}\,(m\,\om_\tsf{A}\,\c Q) \approx   {\rm det}\,\tsf{Q}(\tsf{S}_{\tsf{i}}), \qquad
       L \le |p_{i+1}-p_{i+2}|,   }
where
$\c Q_{ab}= (\tsf{A}/\tsf{M})\, \Sig_{k=1}^\tsf{A}\,\,m_k\,(x_{\a k}-X_\a)\,(x_{\a k}-X_\a)$ is the mass quadrupole--monopole
tensor, $\vec X$ is the center of mass vector,
while   the total energy  equal to $E_{\la \vec p \ra}(L,\tsf{S}_{\tsf{i}})\sS= T + U(\c Q) \approx
  {3\,(s+2)}\,(2\,s)^{-1}\,\om_{\la r\sS=2 \ra}\, \geo{p}{1- \kap(s)}\,{\bar E_i}{^{\kap(s)/3}}$
restates   the formula onto the total energy obtained  here.
The same  formulas (for $r\sS=2$ and $s\sS=2$) have been derived much earlier \cite{cerkSzym} by investigating
a simplified form of a cranked harmonic oscillator model in which the ${\rm U}(3)$ tensor
terms of [2,0,0] and [0,0,-2] type have been neglected in the procedure of diagonalization of the Routhian function:
$H_{\vec \om,\Om}= \sum_a \om_a\,A_{aa}(m) -\Om\,L_z(m)|_{\vec \om=(\om_x,\om_y,\om_z)}$.

\noi For $(L,p,\gam) \in \tsf{M}^+_{\vec{p}} $ one finds:
\begin{list}{}{\itemindent=2mm\parsep=0mm\itemsep=0mm\topsep =0mm}
\item[$(A)$]  $\tsf{P}_-\cap\tsf{S}_{\tsf{3}}\sS=(p_1-p_2,p_3,0)$,
\item[$(B)$]  $\tsf{P}_+\cap\tsf{S}_{\tsf{3}}\sS=(p_2-p_3,p_1,0)$,
\end{list}
and if  $\mu<\lam$ then
\begin{list}{}{\itemindent=2mm\parsep=0mm\itemsep=0mm\topsep =0mm}
\item[$(C)$]  $\tsf{P}_-\cap\tsf{S}_2\sS=(p_1-p_3,p_2,0)$
 and   $\tsf{P}_+\cap \tsf{S}_2\sS=(2\,\sqrt{\lam\,\mu},p_2,0)$,
\end{list} else,
\begin{list}{}{\itemindent=2mm\parsep=0mm\itemsep=0mm\topsep =0mm}
\item[$(D)$]  $\tsf{P}_+\cap\tsf{S}_2\sS=(p_1-p_3,p_2,0)$
 and  $\tsf{P}_-\cap\tsf{S}_2\sS=(2\,\sqrt{\lam\,\mu},p_2,0)$.
\end{list}
\vskip  2mm
%
       %\begin{comment}
\begin{figure}[!b]
  \centering    \includegraphics[scale=.8]{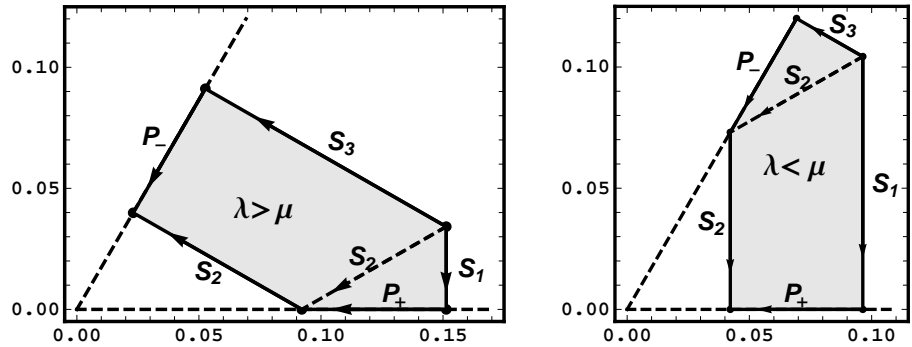}
  \caption{The projection of   $(\lam,\mu)$ states  onto the $(\bt,\Gam)$ plane   }
 \end{figure}
      %\end{comment}
The structure of $\tsf{S}$ and $\tsf{P}$  bands on $[\vec p]\toast$ depends on the sign of expression $\lam -\mu$.

If $\mu < \lam$, then  the   minima  of $H\sS={\rm det}(\tsf{Q})$ form  two bands:
\begin{list}{}{\itemindent=2mm\parsep=0mm\itemsep=0mm\topsep =0mm}
\item[$(a)$]  $\tsf{S}_3$ for $0<L\le\lam$, and,
\item[$(b)$]   $\tsf{P}_-$ for $\lam \le L\le \lam+\mu$ ellipsoids,
\end{list} while  hamiltonian   maxima decouple into three  bands:
\begin{list}{}{\itemindent=2mm\parsep=0mm\itemsep=0mm\topsep =0mm}
\item[$(c)$]  $\tsf{S}_1$ for $0\le L\le\mu$,
\item[$(d)$]  $\tsf{P}_+$ for  $\mu\le L\le 2\,\sqrt{\mu\,\lam}$, and,
\item[$(e)$]  $\tsf{S}_2$ for  $2\,\sqrt{\mu\,\lam}\le L\le \lam +\mu$,
\end{list}
If   $\lam <\mu$   hamiltonian   minima decouple into  three bands:
\begin{list}{}{\itemindent=2mm\parsep=0mm\itemsep=0mm\topsep =0mm}
\item[$(a)$] $\tsf{S}_3$ for  $0\le L\le \lam$,
\item[$(b)$] $\tsf{P}_-$ for  $\lam \le L\le  2\,\sqrt{\mu\,\lam}$, and,
\item[$(c)$]  $\tsf{S}_2$ for  $ 2\,\sqrt{\mu\,\lam}\le L\le \lam+\mu$,
\end{list} while hamiltonian  maxima form   two bands:
\begin{list}{}{\itemindent=2mm\parsep=0mm\itemsep=0mm\topsep =0mm}
\item[$(d)$]   $\tsf{S}_1$ for  $0\le L\le \mu$, and,
\item[$(e)$]  $\tsf{P}_+$ for $\mu\le L\le \lam+\mu$.
\end{list}
In both the cases $\tsf{S}_2$ ellipsoids for  $0\le L \le 2\,\sqrt{\smash[b]{\lam\,\mu}}$ are
vibrationally unstable states.

The graphic representation of   two  structures of equilibrium bands found
for a family of  orbits $(a)$ $\mu<\lam $ and $(b)$ $\lam<\mu$ are studied on  Figure 1.
the $(\lam,\mu)$ states of  ${\rm su}\co(3)$ orbit are projected on
the $(\bt,\Gam)$ plane: $(x,y) = (\bt\,\cos \Gam,\bt \sin \Gam)$.
The parameters $(x,y)$ fix the eigenvalues $(P_1,P_2,P_3)$ according  to  the
rules presented  in eqs.(\ref{P123a}, \ref{P123b}).

The left graphic is obtained  for the values $(\lam,\mu,p_3)=(50,15,100)$  while the right for $(15,50,100)$.
$\tsf{S}_i$  and $\tsf{P}_\pm$ ellipsoids  close the set of $(\lam,\mu)$ states represented by the shadowed area.
If $\mu\le \lam$, then parameter $\Gam$ of  the maximal state $L\sS=\lam+\mu$ is equal to $\Gam\sS=60°$ (prolate ellipsoid) else
$\Gam\sS=0°$ (oblate ellipsoids). On both pictures, the angular momentum of ellipsoids  increases accordingly with the direction
of arrows.
\section{\label{PGAM} Wobbling  motion for Hamiltonian $H = E_0 + \tfrac 1 4\,\om(L)\,Q$.}
% \subsection{Coordinates--momenta  system     $(p,\gam)$ and $(Q,\vth)$}
\subsection{Canonical pair of coordinates  conjugated to  momenta $(L,Q)$}
Coordinates $L,Q$ commute, so  $L,Q\sS=Q_{L}(p,\gam)$ are constants of motion of Hamiltonians $H\sS=H_{\om,r,s=2}$; hence,
$H\sS=H(L,p,\gam)\sS=  h_0 + \tfrac 1 4\,\om(L)\,Q$. Then
$\dot p = -H\ord{0,0,1}(L,p,\gam)= 2\,L^{-2}\,G_{\vec{p}}(p)\,\om(L)\,\cos \gam\,\sin\,\gam$
where according to  the second relation in eq.(\ref{pGmix}) on the surface $Q\sS={\rm const},L\sS={\rm const}$:
 $\gam = \eps_\gam\,\gam^{+}_{L,Q,p}$,  $\eps_\gam={\rm sign}\,\gam$,
 $\gam^{\pm }_{L,Q,p}=-i\,\ln Z^{\pm}_{\vec{p}}(L,Q,p)$ where
\begg{ \label{Z}  Z^\pm_{\vec{p}}(L,Q,p)=
       {G_{\vec{p}} }{^{-1/2}}(p)\,\bigl[ \tfrac L 2 \,\sqrt{\smash[b]{Q-p}} \pm i\,
       \sqrt{  \smash[b]{-V_{\vec{p}}(L,Q,p)}} \bigr], \\
   \label{V}     V_{\vec{p}}(L,Q,p) = - G_{\vec{p}}(p) -  \tfrac 1 4\,(p-Q)\,{L}^{\;\,2}. }
We get
\begg{ \label{dtp}  \dot p  = %  \eps_{\gam}\,\sig_Q\,L^{-1}\,\sqrt{\smash[b]{V_{\vec{p}}(L,Q,p)\,(p-Q)}}\,\om(L)=
    - \eps_{\gam}\,L^{-1}\,\sqrt{\smash[b]{-V_{\vec{p}}(L,Q,p)}}\,\sqrt{\smash[b]{Q-p}}\,\om(L),  \\
 \label{dtg}       \dot \gam  = (\pl_p \gam_{L,Q,p}) \,\dot p =
   \tfrac 1 2\,\cot \gam\;[G_{\vec{p}}\,(p-Q)]^{-1} \,V_{\vec{p}}(\c L\,(p),Q,p)\times \dot p,   }
Physical interpretation  of roots of the polynomial  $p\to V_{\vec{p}}(L,Q,p$) is established by studying
secular equations onto eigenvectors  of the tensor  $\tsf{Q}$:
\beqm{ \label{Pi}
    0 = {\rm det}\,(\tsf{Q} - P\,\mathbf{1} ) = V_{\vec{p}}(L,Q,P) = - \prod_{k\le 3} (P-P_{k,L,\vec p}(Q)). }
\ntheo{ \label{TPi}  Three roots $P_\a = P_{\a,L,\vec p}(Q)$ are real if
 $(L,Q)\in \tsf{D}_{\vec{p}}\sS=\bigcup_{L} [Q_{\rm min}(L),Q_{\rm max}(L)]$
$($see, eq.$(\ref{QintL}))$.
Set $ \tsf{D}_{\vec{p}}$ decouples
into two subsets $\tsf{D}_{\vec{p}}=  \bigcup_{\sig = \pm   }   \tsf{D}^\sig_{\vec{p}}  $ % \tsf{D}^+_{\vec{p}}\cup \tsf{D}^-_{\vec{p}}$
  where
  $\sig \sS\equiv \sig_Q =  {\rm sign}\,(Q-p_2)$.
Roots $\vec P=\vec P(L,Q)$  obey the set of rules  $A_{\sig_Q} [\vec P,Q]$ where
\begg{  A_-[\vec P,Q] =     p_3 \le P_{3} \le  Q \le p_2 \le P_{2}  \le  P_{1} \le p_1, \nonu \\
        A_+[\vec P,Q] =     p_3 \le P_{3} \le p_2 \le P_2  \le  Q  \le   P_{1} \le p_1. \nonu   }}
The set $\c D_{L,\vec p}$ of points $\vec d = (Q,p)$  is selected by a pair of conditions
\begin{list}{}{\itemindent=2mm\parsep=0mm\itemsep=0mm\topsep =0mm}
\item[$(a)$]  $0< \sin^2\,\gam\,\cos^2 \gam\,(G_{\vec{p}}(p)/L)^2 $,
\item[$(b)$] $0 < R_{L,Q,p}{^2}$.
\end{list}
Transforming $\gam \to \gam_{L,Q,p}$ and using eqs.(\ref{pGmix})  we get
\begg{  \c D_{L,\vec p}=\{(Q,p), 0 \le  V_{\vec{p}}(L,Q,p)\times (p-Q),\,0\le G_{\vec{p}}(Q)\,(p-Q)\} =
  \c D^{-}_{\vec{p}}\cup  \c D^{+}_{L,\vec p},\\
  \c D^-_{L,\vec p}\cap  \c D^+_{L,\vec p}\sS=(p_2,p_2), \\
  \vec d \in D^{\sig}_{\vec{p}} \setminus \overline{D^{\sig}_{L,\vec p}} \Rightarrow ( 0<\sig\,\,(d_1-p_2),\,0<-\sig\,\,(d_2-p_2)).  }
From:  $(a)$ formula  (\ref{Pi}), $(b)$ the identity
$V_{\vec{p}}(L,Q,p) = \tfrac 1 4\,L^2\,( Q- Q_{L,\vec p}(p,0))$, $(c)$ theorem \ref{TPi}
and $(d)$ association $(d)$ $A_\pm \leftrightarrow \c D^\pm_{L,\vec p}$ one  finds
\begg{ \c D^{+}_{L, \vec p}= \bigcup_{p_2 \le Q \le Q_{\rm max}(L)} (Q,\Del_+(L,Q)),
    \qquad   \c D^-_{L,\vec p}= \bigcup_{Q_{\rm min}(L)\le Q\le p_2} (Q,\Del_-(L,Q)),\\
  \label{DEL}   \Del_+(L,Q) =    [P_{3,\vec p}(L,Q),P_{2,\vec p}(L,Q)], \\
  \Del_-(L,Q)= [P_{2,\vec p}(L,Q),P_{1,\vec p}(L,Q)].   }
The family of  $\tsf{P}_\pm$  ellipsoids is  selected  assuming
$(Q,p) \sS=(Q_{L,\vec p}(p_\pm (L),p_\pm(L))$ where  $p_\pm(L)\in D^\pm_{L,\vec p}$ and the
assumption $\cos \gam_{L,Q,p}\sS=1$.

The explicit expressions for roots $P_\a\equiv P_{\a,\vec p}(L,Q)$ are determined from the following formulas:
\begg{ \label{P123a} P_{\a,\vec p}(L,Q)  =
       \tfrac 1 3\,S_1 \times\left( 1 +  2\,\bt\,\cos\,(\Gam - \tfrac 2 3\,(\a-1)\,\pi)\right), \\
       \label{P123b} \bt= \frac{S_L}{S_1},\qquad  \Gam = \Gam_L(Q) =
        \frac 1 3\, \arccos\left(-\frac{4\,S^A_{111}+9\,L^2\,(S_1-3\,Q)}{8\,{S_L}^3}\right).    }
$\tsf{S}_{i}$ ellipsoids ($Q\sS=p_i$)  are triaxially  deformed. One finds
\begin{list}{}{\itemindent=2mm\parsep=0mm\itemsep=0mm\topsep =0mm}
\item[$(a)$]   $\vec P=(P_+,P_-,P_0)$ for  $\tsf{S}_3$ and for  $\tsf{S}_2$ ellipsoids if $2\,\sqrt{\lam\,\mu}<L$ and $\lam<\mu$, % ($(Q,p)\in \c D^-_{\vec{p}}$),
\item[$(b)$]   $\vec P=(P_0,P_+,P_-)$ for  $\tsf{S}_1$  and for $\tsf{S}_2$ ellipsoids if $2 \,\sqrt{\lam\,\mu}<L$ and $\lam<\mu$, %($(Q,p)\in \c D^+{\vec{p}}$),
\item[$(c_<)$]   $\vec P=(P_+,P_0,P_-)$  for  $\tsf{S}_2$ ellipsoids  if $L<2\,\sqrt{\lam\,\mu}$,
\item[$(c_>)$]   $\vec P=(P_+,P_0,P_-)$  for  $\tsf{S}_2$ ellipsoids  if $2\,\sqrt{\lam\,\mu}<L$,

\end{list}
where
\beqm{ \label{P123c}  P_0=p_i, \qquad
        P_\pm = \tfrac 1 2\,(p_{i+1} + p_{i+2}) \pm \tfrac 1 2\,\sqrt{\smash[b]{(p_{i+1}- p_{i+2})^2-L^2}}. }
The shapes of  P$_\pm$ ellipsoids  are   axially symmetric:
 \begin{list}{}{\itemindent=2mm\parsep=0mm\itemsep=0mm\topsep =0mm}
\item[$(d)$]  $\vec P_+ = (P_{a+},P_{b+},P_{b+})$,
\item[$(e)$]  $\vec P_- =(P_{b-},P_{b-},P_{a-})$,
\end{list}
where
\beqm{ \label{P123d}  P_{a\mu } = \tfrac 1 3\,(S_1 + \mu\, 2\,S_L), \qquad P_{b\mu }= \tfrac 1 3\,(S_1 - \mu S_L).    }
Solutions  of the equations of motions are determined with the help of elliptic functions.
Let
\begg{ \c D^{\;\sig}_{L,\vec p} \ni   (Q,p)   \to
       I_{\la L,Q\ra }(p)  \equiv   I^{\sig_Q} _{[\vec P,Q]}(p),  \qquad \vec P \equiv \vec P_{\vec{p}}(L,Q),  \\
       I^\sig_{[u]}(x) =- 2\,{C^\sig_{[u]}}{^{-{1/2}}}\,F(\arcsin\,\sqrt{ \smash[b]{A^\sig_{[u]}(x)}}\;|\,B^\sig_{[u]}), \\
  J^\sig_{[u]}(x) =  2\,{C^\sig_{[u]}}{^{-{1/2}}}\,\Pi (D^\sig_{[u]},
\,\arcsin\,\sqrt{ \smash[b]{A^\sig_{[u]}(x)}}\;|\,B^\sig_{[u]}) }
where  $[u]\equiv [a,b,c,d]$. Then
\begg{    A^-_{[u]} (x)= \frac{  (a-d) (x-b)} {(a-b)\,(x-d)}, \quad B^-_{[u]} =\frac{(a-b) (d-c)}{(b-c) (a-d)}, \qquad
     C^-_{[u]} = (a-d)\,(b-c), \\
    A^+_{[u]} (x)= \frac{ (c-d) (b-x)} {(c-b)\,(d-x)},\quad B^+_{[u]}=\frac{(b-c) (a-d)}{(a-b) (d-c)},
      \qquad C^+_{[u]} = (a-b)\,(d-c), \\
    D^-_{[a,b,c,d]} = D^+_{[c,b,a,d]} =  \frac {a-b}{a-d}. }
%      A^+_{[u]} (x)= \frac{ (a-b) (c-x)} {(c-b)\,(a-x)},\quad B^+_{[u]}=\frac{(c-b) (a-d)}{(a-b) (c-d)},
%      \qquad C^+_{[u]} = (a-b)\,(d-c)  }
%
These functions obey the relations $X^+_{[a,b,c,d]}=X^-_{[c,b,a,d]}$. We have
\begin{list}{}{\itemindent=2mm\parsep=0mm\itemsep=0mm\topsep =2mm}
\item[$(a)$]   $A^{-}_{[u]} ([b,a])\sS=[0,1]$ ,  $A^{+}([c,b])\sS=[1,0]$,
\item[$(b)$]  $I^\sig_{[u]}(u_2)= 0, \, (I^\sig_{[u]})\pr (x)=[u,x]_-$,   $(u_4,x)\in \c D^\sig_{\vec{p}}$,
\item[$(c)$]  $J^\sig _{[u]}(u_2)= 0, \, (J^\sig_{[u]})\pr (x) =  [u,x]_+$, $(u_4,x)\in \c D^\sig_{\vec{p}}$,
\item[$(d)$]   % X_{\la L,Q\ra }(x),
               $I_{\la L,Q\ra }(x)  \sS< 0\sS < - \sig_Q\,J_{\la L,Q\ra }(x)$, $X\sS=I,J,\,(Q,x)\in \c D^\sig_{\vec{p}}$,
                % \equiv X^{\sig_Q}_{[\vec P(Q),Q]}$
\item[$(e)$]  $0\le {\eps_X}{^{(1-\sig_Q)/2}}\, {X_{\la L,Q\ra }}\pr(x)$,
  $\eps_J=-\eps_I=1$, $(Q,x)\in \c D^\sig_{\vec{p}}$,
\item[$(f)$] $d\gam^{\pm}_{L,Q,p}\sS = \mp ( \tfrac 1 2\,[L,Q,p]_+ \times  dL  +
              \tfrac 1  4\, L\,[L,Q,x]_-\times dQ)  + \gam^{\pm}_{,p} \times dp $,
\item[$(g)$] $(X^\sig_{[u]} \circ \hat X^{\sig}_{[u]})(y)=y$, $X\sS=L,J$,
\item[$(h)$] $\hat X^{\sig}_{[u]}(y+2\,k\,X^{\sig}_{[u]}(u_{2-\sig}))= \hat X^{\sig}_{[u]}(y)$,
      $k=\pm 1,\pm 2,\ldots$,
\item[$(i)$]   $0 \sS < - (-\eps_X){^{ (1-\sig)/2 }}\,X^{\sig}_{[u]}(u_{2-\sig}) =
        2\,(C^\sig_{[u]})^{-1/2} \times X(B^{\sig} _{[u]})$,
\item[$(j)$] $F( \pi/2|\;m)  \equiv K(m)$,   $ \Pi(a,\pi/2|\,m) \equiv \Pi(a|\,m)$.
 \end{list}
where
\begg{ \label{dgam}   [u,x]_\pm=  (u_4-x)^{\pm 1/2}\times [\Pi _{k\le 3}(x - u_i)]^{-1/2}, \intertext{and}
        \hat I^{\sig}_{[u]}(y) = \hat A^{\sig}_{[u]}
         \circ {\rm SN}^2 (\,\tfrac 1 2\,(C^\sig_{[u]})^{1/2}\,y,B^\sig_{[u]}\,),\\
      \hat A^-_{[a,b,c,d]}(x) = \frac{ c\,(a - d) + a\,(d -c)\,x}{ a - d + (d-c)\,x},
     \quad A^{+}_{[a,b,c,d]}(x)= A^-_{[b,c,d,a]}(x),   }
$[L,Q,x]_\pm \equiv [u(L,Q),x]_\pm$ while
$F(a|m)$ and $\Pi(a,b|m))$ ($K(m)$ and  $\Pi(a,m)$) represent   elliptic incomplete (complete) integrals of first and third  kind,
respectively. ${\rm SN} (u,m)\sS=\sin\circ {\rm am}(u,m)$ where ${\rm am}\,(u,m)\!: {\rm am}(F(\vp,m),m)=\vp$ is
the Jacobi amplitude.
%
%
% \ndef{ N_{\vec{p}}(k)= \{ (L,Q,\vp),}
%\,   \vth\in [(2\,k-1)\,\Del_{\la L,Q\ra},\,(2\,k+1)\,\Del_{\la L,Q\ra}],\,
%
\ndef{ $\tsf{N}_{\vec{p}}= \{ [(L,Q),\,(\psi,\,{\rm Mod}\,(2\,\pi),\vth\,{\rm Mod}\,(2\,\Del_{\la L,Q\ra}))],
     (L,Q) \in \tsf{D}_{\vec{p}}, \Om_{\tsf{N}_{\vec{p}}(k)} = dL \we d\psi + dQ \we d\vth \}$ where
\beqm{     \Del_{\la L,Q\ra} =-\tfrac 1 4\,L\,I_{\la L,Q\ra}(p_{2-\sig}) =
           \tfrac 1 2 \,L\,{C_{\la L,Q\ra}}{^{-1/2}}\,K(B_{\la L,Q\ra}).    }}
Here and at points $(d,e)$ we simplified the notation $X_{\la L,Q \ra} \equiv X^{\sig_Q}_{[\vec P_{\vec{p}}(L,Q),Q]}$.
\ntheo{ Let
\beqm{ \label{volQ} \c V_{\vec p,L}= \c V_{\vec p,L}(Q^L_{\rm max}),\qquad   \c V_{\vec p,L}(Q)=
      \frac 2 {\pi}\,\int^Q_{Q^L_{\rm min}} dt\,\Del_{[L,t]},  }
where, $Q^L_{\rm min(max)}$ are  functions of $q\equiv (\vec p,L)$ defined in eq.$(\ref{QintL})$.
 For integer values  $\lam=p_1-p_2,\,\mu\sS=p_2-p_3$ and $L$
volume integral $\c V_{\vec p,L}$ takes integer values determined from the formula }
 \begg{ \label{volF} \c V_{\vec p,L} = \beg{cases}{ L   & {\rm if~} L \in [0,{\rm min}(\lam,\mu)], \\
          {\rm min}(\lam,\mu)  &  {\rm if~}L \in [ {\rm min}(\lam,\mu),{\rm max}(\lam,\mu)], \\
          \lam +\mu - L  &     {\rm if~} L \in [  {\rm max}(\lam,\mu),\lam +\mu].}}   }
Proof of this formula has been  verified by  performing the numerical integration of integral in eq.(\ref{volQ}).
The independent proof  will be performed in the  Sec. VI  [see,
the comments  between the  formulas  (\ref{Vol},\ref{delD},\ref{delF})], where
the quantity $\c V_{\vec p,L}$ is considered as a classic limit for  quantum
coefficients determining ${\rm u}(3)\to {\rm so}(3)$ reduction of irreducible unitary representation
(IUR):   $[\vec p\,]$.
\ntheo{ Mapping $\vk\!: M_{\vec{p}} \ni [(L\pr,p),(\vp,\gam)]\to  n= [ (L,Q),(\psi,\vth)]\in \tsf{N}_{\vec{p}}(k)$
established by the relations
\begg{ \label{kap1}  L= L\pr, \qquad Q=Q_L(p,\gam)= p+4\,L^{-2}\,G_{\vec{p}}(p)\,\cos^2\gam,\\
       \label{kap2} \vth =   \tfrac 1 4\,\eps_\gam\,L\,I_{\la L,Q_L(p,\gam)\ra }(p)\, {\rm Mod}\,( 2\,\Del_{\la L,Q_L(p,\gam)\ra}),   \\
       \label{kap3}  \psi =  \vp  + \tfrac 1 2\,\eps_\gam \,J_{\la L,Q_L(p,\gam)\ra }(p),}
defines  canonical isomorphism: $\Om_{\tsf{M}_{\vec{p}}}(m) \sS= (\vk \co\,\Om_{\tsf{N}_{\vec{p}}})(m)\sS=dL\pr\we d\vp  + dp\we d\gam$.
Inversion $\vk^{-1}$ follows from the relations
\begg{  \label{ikap1}   p=p_{L,Q}(\vth),     \quad
      \vp=   \psi - \tfrac 1 2\,\eps_{\vth}\,J_{\la L,Q\ra}\circ p_{L,Q}(\vth), \qquad
      \gam=   -i\,\ln Z^{-\eps_\vth} _{\vec{p}}(L,Q,p_{L,Q}(\vth)), \\
  \label{ikap2}   p_{L,Q}(\vth) = \hat I_{\la L,Q\ra}( 4\,\eps_{\vth}\,L^{-1}\,\vth)=
    \hat A_{\la L,Q \ra}\circ {\rm SN}(\tfrac {\eps_{\vth}}  2\,L^{-1}\,{C_{\la L,Q\ra}}{^{1/2}}\,\vth,  B_{\la L,Q\ra}),\\
   \label{ikap3}  \eps_\vth={\rm \sign}\,\sin\,(\pi/\Del_{\la L,Q\ra}\,\vth). }
$p_{L,Q}(\vth) $ is smooth periodic function:   $p_{L,Q}(\vth+2\,\Del_{\la L,Q\ra})\sS=p_{L,Q}(\vth)$.  }
{\em Proof}. The proof of the theorem results  from the construction of a pair of generating functions $S_\pm$:
\[ L\pr\,d\vp + p\,d\gam- L\,d\psi - Q\,d\vth = d\,[p\,\gam - Q\,\vth - L\,\psi +   S_\pm(\vp,L,Q,p)],  \]
 where
$ S_\pm(L,Q|p,\vp) = \vp\,L + \bar S_\pm(L,Q,p)$ and
\begg{  \bar S_{\eps_\gam}   (L,Q,p)   = \beg{cases}{- {\eps_\gam} \,\int_{p_2}^p \gam^+_{L,Q,t} \times dt &
          \text{if~} (Q,p)\in \c D^-_{\vec{p}}, \\
         \hph{-} {\eps_\gam} \, \int^{p_2}_{p}    \gam^+_{L,Q,t} \times dt  & \text{if~}   (Q,p)\in \c D^+_{\vec{p}}.}}
Here,  $\gam^+_{x,y,z} \sS= -i\,\ln\,Z^{+}_{\vec{p}}(x,y,z)$. From the definitions of $S_{\eps_\gam}$ and
points $(e,f,i)$ we have $(dS_{\eps_\gam})(\vp,L,Q,p)=L\pr\,d\vp-\gam\,dp+\psi\,dL +\vth\,dQ$; hence
considering point $(f)$ and eqs.(\ref{dgam}),
\begg{  \gam = - S_{\eps_\gam,p} = \eps_{\gam}\,\gam^+_{L,Q,p}, \qquad L\pr = S_{\eps_\gam,\vp}=L  ,\\
        \vth =   S_{\eps_\gam,Q} =  \tfrac {\eps_\gam}  4\,L\,I_{\la L\pr,Q\ra }(p),  \qquad
        \psi =  S_{\eps_\gam,L} =  \vp + \tfrac {\eps_\gam} 2\,J _{\la L\pr,Q\ra }(p).   }
which restate formulas in eqs.(\ref{kap1}--\ref{kap2}), as well with the help of them
the proof  of the pull back rule: $(\kap\co \Om_{\tsf{N}_{\vec{p}}})(m)=\Om_{\tsf{M}_{\vec{p}}}(m)$ is turning into
the well--known identity relation.

The  periodicity of functions $\vth \to g(\vth)$ for $g=\vp,\gam$ follows from
periodicity of $p_{L,Q}(\vth)$ and periodicity of expression onto signature $\eps_\vth$.

In order to prove the consistence  of definition for $\eps_\vth$ (see, eq.(\ref{ikap2}) with  formula (\ref{kap2})),
we should point that the  condition $\gam=0$ follows from  the requirement $V_{\vec{p}}(L,Q,p)=0$; hence,
  $L={\rm const},\, Q\sS={\rm const},\,\vp={\rm const}$ define a pair of smooth curves $\Upsilon^\pm_{L,Q}$:
$\Del_{\sig_Q}(L,Q) \ni p \to v_\pm (p) =[(p,\gam^{\pm}_{L,Q,p}),\vth_{L,Q}(p) ]$ which cross the
points $v_\pm(i),\,i=1,2$  that  $\{p(1),p(2)\} = \pl \Del_{\sig_ Q}(L,Q)$. We have
\beqm {    v_{\pm}(1) =  [(P_{2,\vec p}(L,Q), 0),0], \qquad
           v_{\pm}(2) =  [(P_{2-\sig_Q,\vec p}(L, Q),0),\mp \Del_{\la L,Q\ra}],   }
Since $\vth$ is a periodic variable   $\vth \equiv \vth\, {\rm Mod}(2\,\Del_{\la L,Q\ra})$, hence:
\begin{list}{}{\itemindent=2mm\parsep=0mm\itemsep=0mm\topsep =0mm}
  \item[$(a)$]    $\vth$ is the parameter of the curve
 $\Upsilon_{\la L,Q\ra}= \Upsilon^+_{\la L,Q\ra}\cup \Upsilon^-_{\la L,Q\ra}$ homotopic
  to a circle,
\item[ $(b)$] for points $\Gam_{\la L,Q\ra}\setminus \{v(1),v(2)\}$: $\eps_\gam\,\eps_\vth=-1$.
\end{list}
Proof of formulas (\ref{ikap1},\ref{ikap2}) results  trivially from application of the rule at point $(b)$.
\subsection{Frequency of periodic  motion }
Since coordinates $ (L,Q,\psi,\vth) = \kap^{-1}(L,p,\vp,\gam)$ are canonical,
the equations   of motion take  the form:
\beqm{  \dot{\vth}  = \tfrac 1 4\,\om(L), \qquad  \dot \psi = \tfrac 1 4\,Q\,\om\pr(L),
     \qquad \dot Q=\dot L=0;   }
hence $q(t) = \dot q(L,Q)\,t+q_0$, $q\sS=\psi,\vth$. Let
\begg{        \Om_{\vth}  = 2\,\pi\,{T_\vth}{^{-1}}  =
           \frac{\pi}{2} \times \frac{  (C^\sig_{\la L,Q\ra})^{1/2} \,\om(L) } {L\,K(B^\sig_{\la L,Q\ra}) }, }
where is the frequency associated with the period of time $T_{\vth} = 2\,\Del_{\la L,Q\ra}/\dot{\vth}$ for
the  mode $(Q,\vth)$.
The  states selected  at the point $(c_>)$  of the list including   equation (\ref{P123c}) represent
the saddle point.    We have:
$B^\sig{[\vec P(p_2),p_2]}= 1$,  $K(B^\sig _{[\vec P(L,p_2),p_2]})=\infty$,
$C^\sig _{[\vec P(L,p_2),p_2]}=-C({\rm S}_2,L)<0$ (see, eq.(\ref{CS1}));   hence,  $\Om_\vth = 0$.
For    remaining   points ($(a,b,c_<,d,e)$) of this list   $B^\sig_{\la L,Q\ra}=0$;
hence, $K(B^\sig_{\la L,Q\ra})\sS=K(0)= \tfrac 1 2\,\pi$,

\begg { \label{Omwobb}  \Om_\vth(\tsf{X},\vec p) = L^{-1}\,{C_{\vec{p}}}{^{1/2}}(\tsf{X},L)\,\om(L) \approx
        \geo{p}{-2}\,L\,{C_{\vec{p}}}{^{1/2}}(\tsf{X},L)\,\om_{\la r \ra},
\intertext{where }
  \label{CS1}   C_{\vec{p}}(\tsf{S}_{\tsf i},L) =   (p_{i+1}-p_i)\,(p_{i+2}-p_i) + \tfrac 1 4\,L^2 \equiv
      (P_{i+1}- P_{i})\,(P_{i-1}- P_{i}), \\
   \label{CS2}  C_{\vec{p}}(\tsf{P}_\pm,L) =
   (4/27)\,L^{-2}\times [(\pm S^A_{111}+2\,{S_L}{^3})\,S_L  + \tfrac 1 3\,L^2\,{S_L}^2].        }
\subsection{Coalgebras of   the wobbling mode in collective dynamics}
The natural physical interpretation of the mode spanned  by the coordinates $(Q,\vth)$ is found by considering the
 following formulas:
\def\coo{ {^{\ast\!}} }
\Meq{to*L}{
\begg{   \coo\c L_{\a}= \c L_{\vec p,L,Q, P_{\a,\vec p}(L,Q))}\circ p_{L,Q}(\vth), \\
    \label{tocLi2}  {\c L_{\vec p,L,Q,P}} {^2}(p) = \frac 3 4\, \frac {p-P}{p-Q} \times
          \frac{ G_{\vec{p}}(Q)}{G_{\vec{p}}(P) } \times \frac {L^2}{  (3\,P-S_{1})^2 - {S_L}{^2} }, }}
where  $\coo\vec{\c L\;}=(\coo\c L_1,\coo\c L_2,\coo\c L_3)$  represent the  components of angular momentum vector  in the body frame
 of references (${^\ast}$BF); the function $F_{p,L}(x,\gam)$ has been defined  in eq.(\ref{toF}) and
$P_\a=P_{\a,\vec p}(L,Q))$ are eigenvalues determined  in eq.(\ref{P123a}).

In order to explain  the relations  between  ${^\ast}$BF and  two  reference  frames discussed earlier, let
us consider the list of the references frames
\llist{ \itt{a}  the inertial frame (IF):  $(\bo{a}^1,\bo a^2) \Rightarrow (
  \vec l(\tsf{A}) = \vec{\tsf{L}},\,\tsf {q} (\tsf{A}) = \tsf{Q})$,
\itt{b} the  angular momentum frame (AMF): $(\bo{b}^1,\bo b^2, \tsf {L}) \Rightarrow (\vec l(\tsf{B}) = (0,0, |\vec{\tsf{L}}|),
  \tsf{q}(\tsf{B})  \sS= Q)$ and
\itt{c}   the body frame  ($\coo$BF): $(\coo\bo{c}^1,\coo \bo c^2,\tsf{L}, \vp_L) \Rightarrow
      (\coo\vec{\c L\;} = \vec l( \coo \tsf{C}),  \tsf{q}(\coo \tsf{C}) = {\rm diag}\,(P_1,P_2,P_3)$). }
%
% \end{document}
where $ l_a (\tsf{X}) = - i\,\Sig_{bc} \eps_{abc}\, (X_{bc}-X_{cb})$, $\tsf{q}(\tsf{X}) = \tfrac 1 2\,(\tsf{X} + \trx 1\,\tsf{X})$,
 $\tsf{X}\sS=\tsf{A,B,\coo C}$ and the   following diagram:
\begg{  \label{ILB}   \beg{CD}{
   \tsf{A}   @  <\hph{xx}  { \ds {\rm Ad}^\ast_{\ds \car{\tsf{L}}{\;} }}   \hph{xx}   <<
    \tsf{B}   @ < \hph{x}   { \ds {\rm Ad}^\ast_{\ds R_3(\coo\vp_L)   \cdot \carr{\c L\;}{\!\!\cov}{\coo\!}} } \hph{x}<< \coo\tsf{C} \\
   @ A  \bo a^2 \cdot \,\trx 1\bo a^1 AA  @  A  \bo b^2 \cdot \,\trx 1\bo b^1   AA  @ A \bo c^2 \cdot \,\trx 1\bo c^1  AA  \\
% } }\end{document}
%
   \bo a^u @ << \hph{xxxx}  {\textstyle \car{\,\tsf  L }{} } \hph{xxxx} < \bo b^u
           @ << \hph{xxx} {\textstyle  R_3(\coo\vp_L) \cdot \carr{\c L\;}{\!\!\cov}{\coo\!} } \hph{xxx} < \coo\bo c^u,  }}
where   $\bo x^u= (\vec x^u_1,\ldots,\vec x^u_n),\,u\sS=1,2,\,\bo x=\bo a,\bo b,\bo c$ and
${\rm Ad}^\ast_{x}(y)=x\cdot y \cdot y^{-1}$.

Formula  (\ref{tocLi2}) has been derived  using the equation:
${\coo{\c L}_{\a}}{^2}  = [{u_3}^{2}(P_\a) + \Sig_{i\le 2}(Q_{3-i}\,q_{3-i}-q_i\,q_3)^2]^{-1}\times {u_3}^{2}(P_\a) $,
$ {u_3}^{2}(x) = (Q_1- x)\,(Q_2-x)-{q_3}{^2}$,
definitions  (\ref{pGmix})  and the identity relations
$Q= Q_L(P,\gam=0)$,   $ L^6\,G_{\vec{p}}(Q)  = - G_{\vec{p}}(P)\,F_{\vec p,L}(P,\gam\sS=0)$.

Formula (\ref{tocLi2}) says that the canonical pair $(Q,\vth)$  represents the nonlinear
model of the nuclear wobbling motion.\cite{BohrMII,Mar,Naz}  If  $X\sS={\rm S}_3$ or $X\sS={\rm P}_-$,
then  $\Del E\sS \approx  \hbar\,\Om_\vth({\rm X},\vec p)$ is positive
and it determines the  one--bozon energy excitation within  the harmonic approximation of vibrational expansion
of equations of motion.

In some future paper  we want to discuss the presented  approach
as  an  effective model  of {\em restricted  dynamics}   obtained  by studying
   a  collective motion on the coadjoint orbits of the ${\rm Sp}(6,R)$ group.
 The ${\rm U}(3)$   Hamiltonian is generate considering  many particle systems
bounded by a   simple class  of collective potentials: $U = \a(\bt,p_1\,p_2\,p_3)\ {\rm det}^\bt(\c Q)$
 and  restricting      ${\rm Sp}(6,R)$ orbits  to ${\rm U}(3)$:   $\la p_{1},p_{2},p_{3}\ra \to [p_1,p_2,p_3]$.
The  physical effects following from the violation of ${\rm U}(3)$ orbit structure
 can be neglected  if the model  is applied  to  states of $[\vec p]$ orbits  that   $p_{i}-p_{ i+1}\ll p_{3}$.

Even if the wobbling motion  is treated in the limit of small amplitude vibration
and it is studied for the simplest type of equilibrium  bands
 formed  by $\tsf{S}_{\tsf i}$ ellipsoids,   ${\rm Sp}(6,R)$ approach
requires  much more advanced tools.
This approach has    to employ the six--dimensional phase space\cite{cerkJMP} spanned
by so--called odd--parity--signature phase space coordinates\cite{cerkJMP}\!:
$\Om_6= \Sig_{\a}\; d\bar{ \tsf{p}}_{\a}\we d\bar {\vp}_\a$, $(\bar{\tsf{p}}_1,\bar{\tsf{ p}}_2,
 \bar{\tsf {p}}_3) \in \c P(\vec c),\, \bar \vp_\a= \bar \vp_\a \,{\rm Mod}\,(2\,\pi)$,
$\vec c=(p_{1},p_{2},p_{3},L)$.
In the vibrational limit $\c P(\vec c) \mapsto \Re^3_{+},\,$ and physical interpretation  of
small amplitude vibrations reduces to   the  discussion of  coefficients
 $\gam_{\a}(\vec p, {\rm i})$. They are found by  considering  the following formulas:
\[  \la {L_1}^2 + {L_2}^2 \ra \approx  \hbar\; \Sig_{\a}\,\gam_\a(\vec p; {\rm i})\,{\tsf p}_\a, \qquad
     H^{\rm odd}_{\rm vib}=  \bar H^{\rm odd}_{\rm vib}\circ \rho(q)=
      \Sig_{\a\le 3}\,\om_\a({\rm i}, \vec p)\,{\tsf p}_\a     \]
where
$\rho\!:  (\bar {\tsf p} _1,\ldots, \bar \vp_3) \sS=\rho( {\tsf p}_1,\ldots, { \vp}_3) $ is
a canonical transformation.
We have: \begin{list}{}{\itemindent=2mm\parsep=0mm\itemsep=0mm\topsep =0mm}
\item[$(a)$]   $\lim_{L\to 0} \om({\rm i},\vec p,L)/\Om_{\vth}({\rm S}_{\rm i},\vec p,L)\approx 1$,
\item[$(b)$]  if $p_2 \sS= p_{3} \neq p_1  \vee p_2 \sS= p_1 \neq p_3$  then
 $[(\tsf p_1,\vp_1),\ldots,(\tsf p_3,\vp)] \to [(\tsf p_2,\vp_2),(\tsf p,\tsf \vp_3)],\,\gam_i(\vec p,{\rm i})>0$,
\item[(c)] $\lim_{L\to 0}\lim_{p_2 \to p_i} \rho(q),\,i\sS=1,3$ does not exist.
\end{list}
where   points $(b,c)$ lead to  the following conclusions:
\llist{ \itt{1}  if $p_2\sS=p_{1(3)}\neq p_{3(1)}$ then low energy mode $\om_1({\rm i},\vec p,L)$ is the Goldstone mode,
        \itt{2}  $\lim_{L\to 0}\lim_{p_2 \to p_i} \gam_\a(\vec c; {\rm i})$ does not exist
        \itt{3}  the relation  $\gam_1(\vec p; {\rm i})\gg  \gam_k(\vec p; {\rm i}),\,k\sS=2,3$ establishing the bridge
        between a wobbling motion and the excitation energy:  $\hbar\;\Om_{\vth}({\rm S}_{\rm i},\vec p,L)$
        is not valid, in general.    }
With the reasons  considered  at  the point (2),
 the physical  range  of validity of the  inequality  given at the point $(3)$ reduces
 to some small interval  of low values of angular momentum.
 ${\rm Sp}(6,\Re)$ approach should  be applied  to  the estimation of the upper limit of  this  interval.

It is interesting to compare the diagram  (\ref{ILB}) of ${\rm U}(3)$ collective dynamics
with  a  similar diagram  applied  considering  other  collective models, such as:
\llist{ \itt{1} Dirichlet-- Dedekind--Riemann fluid dynamics,\cite{Chandra}
        \itt{2} ${\rm Sp}(6,\Re)$ dynamical model\cite{cerkJMP} and
         ${\rm SO}(N- 1)$ invariant particle dynamics,\cite{cerkAn}
        \itt{3}   Unified  Bohr collective model.\cite{BohrMII}    }
The algebraic structure of these three models is obtained using the coalgebra $(\tsf{L},\c Q)$.
The latter have been introduced at the end of the section II by the redefinition
of structural constant $\tsf{G}$:   $\tsf{Q} \equiv \tsf  Q_{\tsc {G} = 1}\mapsto \tsf  Q_{\tsc {G} =0}\equiv \c Q$.

In the case  of  $(\tsf{L},\c Q)$ coalgebra, the construction of the body frame  of references (BF) bases on
 mutually commuting  components of tenor $\c Q$:  $\c Q_{ab} =\c Q_{ab}(\bo a^1,\bo a^2)\sS=
     - \tfrac 1 2\,[ ( \bo a^1 -\bo a^2) \cdot \trx 1  (\bo a^1- \bo a^2)]_{ab} +
   \tfrac 1 2\,\Sig_{k\le \tsf{A} } (a^{1}_{ka}- a^2_{ka})\, \Sig_{k\le \tsf{A} } (a^{1}_{kb}- a^2_{kb}) $.
Thus, the  modification
$\coo\,{\rm BF} \to {\rm BF}$ is obtained  as the modification of  the scheme (\ref{ILB}) resulting
from replacements:
\llist{
\itt{1}  $Q \to \c Q$,   $\;(\coo \bo c^1,\coo \bo c^2) \to (\bo c^1,\bo c^2)$,
        $\coo\!\vec{\c L\;}\to \vec{\c L\;}$,   ${\rm diag}\,(P_1,P_2,P_3) \mapsto \bo \lam^2$ where
       %  $\coo \tsf{C} \to  \tsf{C}=(\vec{\c L\;},  {\rm diag}(\bo \lam^2))$,
\itt{2} $\bo \lam^2= \c Q(\bo c^1,\bo c^2)$,   $\bo \lam={\rm diag}\,   ({\lam_1},{\lam_2},{\lam_3})$, $\lam_\a>0$,
 \itt{3}  $\vec{\c {L}\;} = \vec{\c {L}\;}(\bo c^1,\bo c^2)=   - i\,[ \Sig_{k\le  \tsf{A}} \vec{c}^{\;1}_k \times \vec{c}^{\;2}_k -
   \Sig_{ k \le  \tsf{A}} \vec{c}^{\;1}_k \times \Sig_{k\le  \tsf{A}}  \vec{c}^{\;2}_k ]$, }
Since $x\in (\tsf{L},\c Q)$:
 $\{x, {\rm Tr}\,\c Q^{\;k} \}=\{x, {\rm Tr}\,\bo \lam^{2\,k} \}\sS=0$,
 so, the   eigenvalues ${\lam_\a}^{2}$  ( $\lam_\a>0$) are  $(\tsf{L},\c Q)$ invariant functions,
\[   \{\lam_\a, x\} =0,  \qquad  x\in (\vec{\tsf{L}},\c Q);     \]
hence if $\lam_\a\sS\neq  \lam_\bt,\,\a\sS\neq \bt$ then  the orbits of  $(\vec{\tsf{L}},\c Q)$ are six dimensional spaces
$\la \bo \lam \ra = \{ (\vec L,\c Q),  \vec{\tsf{L}} = d \cdot \vec {\c L\;},\,
    \c Q= {\rm Ad}^\ast_{d} ({\rm diag}(\bo \lam)), d \in {\rm SO}(3) \} $.  From   (\ref{ILB}) we get
\beqm{ \label{tod}  d   =  \car{\,\tsf{L}}{\,}\cdot R_3(\vp_L) \cdot \car{\c L\;}{\!\cov},  \qquad
        |\vec{\c L\;}|= |\vec{\tsf{L}}|.  }
and if
\begg{ \label{RB1} \{\tsf{L}_a,\tsf{L}_b\} = \eps_{abc}\,\tsf{L}_c,\qquad
       \{\tsf L_a,\c L_{\a} \} = 0, \qquad         \{\vp_L,{\tsf L}_a \} = \psi_{a}(\vec{\tsf L\;},L_3),
        \quad   \\
       \{\c L_\a, \c L_{\bt} \} = - \eps_{\a\bt\gam}\,\c L_\gam,  \qquad
         \qquad     \{\vp_L,{\c L}_\a \} = \psi_{\a}(\vec{\c L\;},L_3),   }
where $\vec \psi(\vec x,y)$ is given in  eq.(\ref{topsi}) of the  Appendix A,  then
the matrix $d$ obeys the  ${\rm T}\co{\rm SO}(3)$ rules:
$\{\Sig_a u_a\,\tsf{L}_a + \Sig_\a v_\a\,\c L_\a + \Sig_{\a a}\, r_{a\a}\,
  d_{a\a},  d_{b\bt}\}=\Sig_{a \a }\,\Sig_{c \gam}\,
   ( u_a\,\eps_{abc}\,\del_{\a\gam} +v_{\a}\eps_{\a\bt\gam}\del_{ac})\,d_{c\gam }$.

In  the case of  ${\rm u}\co(3)$ coalgebra,  the set of constants  $\lam_\a $
is  replaced by eigenvalues $P_\a$ of tensor $Q$ which
are functions on orbits $[\vec p]$. For the parametrization $(L,Q,\psi,\vth)$ they are
functions of momenta $(L,Q)$ (see, eqs.(\ref{P123a},\ref{P123b})).
In the case of BF parametrization  $(L,Q,\psi,\vth)$  are replaced by $(\coo\vec{\c L\;},\coo\vp_L)$.

The functions  $P_\a\sS=P_\a(\coo\vec{\c L\;})$  define the inversion of mapping $\vec P \mapsto \coo \vec{\c L\;}$
given in
eqs.(\ref{to*L}$a$,\\     \ref{to*L}$b$). In order to get them, let us again consider
the ${\rm u}\co(3)$ Casimir surface. From the first two relations   $S_{1} =  \Sig_{\a}\,P_\a$,
$S_{11}=  \Sig_{\a<\bt} P_\a\,P_\bt - \tfrac 1 4\,\Sig_{\a}\coo{\c L_\a}{^{\;2}}$ we get
 $P_\a = \tfrac 1 2\, ( (-1)^\a\,\c P_L(P)  - P + S_1),\,\a\sS=1,2,\,P\equiv P_3$ for $\c P_L(P)$ given by  eq.(\ref{cP}).
 Rewriting  the third Casimir relation  in the form:
$S_{111}= C_{111} =  P_1\,P_2\,P_3 - \tfrac 1 4\,\Sig_{\a} P_\a\,\coo{\c L_\a}{^{\;2}}$ we get
\begg{  \label{toPL}  0 =\c F_{\vec p,\vec L,r}(P), \\  \c F_{\vec p,\vec L,r}(P)=
   \label{toFL}   [8\,G_{\vec{p}}(P) + (\coo{\c L_1}{^2} + \coo{\c L_2}{^2})\,(3\,P-S_1)]^r  -
    (\coo {\c L_1}{^2} -\coo {\c L_2}{^2})^r\,{\c P_L}{^r}(P),\\
  \label{cP}       \c P_L(P)  = 3^{-1/2}\,\sqrt{4\,{S_L}^{2}-(3\,P - S_1)^2},  }
Since the  roots $P$  are unknown algebraic functions, neither
the explicit analytic form of the mapping: $\coo u = (\coo\c L_1,\coo\c L_2,\coo \c L_3,\coo \vp_L)\mapsto (L,Q,\psi,\vth)$
nor the closed  form of the  Poisson rules for  the  coalgebra  of coordinates $u$ is found.

The separation of degrees of freedom onto the rotational one represented by six  canonical variables
\[   M_{\rm rot}=  (p_M,p_L,p_K,\vp_M,\vp_L,\vp_K),  \qquad 0\le p_X\le 2\, p_L \equiv |L_3|, \; X\sS=M,K \]
and the vibrational one represented by $(\lam_1,\lam_2,\lam_3)$ is possible only for models
employing $(\vec{\tsf L},\c Q)$ coalgebra.
Mapping $M_{\rm rot} \mapsto (\vec{\tsf{L}},\vec{\c L\;},d)\in {\rm T}\co{\rm SO}(3)$ is obtained  using
the formulas  (\ref{tod},\ref{Lxyz},\ref{L123}).
{\em On the orbits of ${\rm u}\co(3)$ coalgebra the separation of degrees of freedom onto the rotational and vibrational
one is violated.}

Even if   we deal with the models selected at the points $(1,2,3)$ where the collective coordinates
$u=(\c L_1,\c L_2,\c L_3,\vp_L$) related with $(\tsf{L},\c Q)$ coalgebra play the essential role,
in the number of physical applications the algebra of coordinates $(\bo b^1,\bo b^2,\tsf{L})$
based on AMF is frequently   more useful than the set of BF coordinates:  $(\bo c^1,\bo c^2,\tsf{L},\vp_L)$.

Firstly,  the range of maps for BF coordinates is  constrained to the points: $\lam_\a\neq \lam_\bt,\,\a\neq \bt$.
Contrary the coordinates  $\vp_L$ is not well established.

Secondly,  since  the mapping $(\bo a^1,\bo a^2) \mapsto (\bo b^1,\bo b^2,\tsf{L})$ is elementary reversible,
the unambiguity group (gauge group) ${\tsf G}_{\rm AMF}$  of AMF  coordinates  is trivial ${\rm G}_{\rm AMF}\sS={\rm id}$.
The BF coordinates:  $(\bo c^1,\bo c^2,\tsf{L},\vp_L)$  have to be considered as the coset space of   equivalent points
determined from   the rules:
\[  (\bo c^1,\bo c^2,\vec{ \tsf{L} },\vp_L) \equiv
(g\cdot   \bo c^1,g\cdot \bo c^2, \vp_L+\Del\vp ),\quad  {R_{3}}{^{-1}}(\Del \vp)\sS=[ (g\cdot \vec{\c L\;})_{\ast}] \cdot g\cdot \car{\c L\;}{}, \]
where $\vec{\c L\;}\sS= \vec{\c L\;}(\bo c^1,\bo c^2)$,
$g\in \tsf{G}_{\rm BF}$. $ \tsf{G}_{\rm BF}$ is the octahedron  symmetry group.
 The construction of Hilbert space for models ${\tsf G}\neq {\rm id}$ requires
 the consideration of additional conditions.
\section{Bohr Somerfield's quantization of momentum Q }
Let
\beqm{ \label{Lxyz}  (\tsf{L}_x, \tsf{L}_y)= \sqrt{p_M\,(2\,|L_3|-p_M)}\times  (\cos\,\vp_M,-\sin\,\vp_M), \qquad \tsf{L}_z=|L_3|- p_M.   }
be  the extension of parametrization of $M\sS=[\vec p]\toast $
onto ${\rm so}(3)$ degrees of freedom. Calculation of the volume integral gives
$\c V_{\vec{p}} = (2\,\pi)^{-3}\,\iint \ldots \int_{M} dp_M\,\we d\vp_M \we dL\we d\psi \we dQ\we d\vth$;  hence,
\beqm{ \label{Vol} \c V_{\vec{p}} = 2\, (2\,\pi)^{-1}\,\iiint  dp_M\we  dL \we dQ\,\Del_{\la L,Q\ra} =
  \int_0^{\lam +\mu} \c V_{\vec{p}}(L,Q)^L_{\rm max}) = \tfrac 1 2\,\lam\,\mu\,(\lam+\mu),}
Physical  interpretation of $\c V_{\vec p,L}(Q_{\rm max})$ is reached  by  considering
the coefficients $d_{[\vec p],L}$ of expansion for branching rules of  ${\rm u}(3) \to {\rm so}(3)$ algebra
reduction:
\[  [\vec p]= \Sig_{L}\,d_{[\vec p],L}\,(L),  \]
for IUR   $[\vec p]$ of  ${\rm u}(3)$ onto IUR  $(L)$  of ${\rm so}(3)$ algebra.
Let
\begg{ \label{delD}  \del_{\vec p,L} =  \c V_{\vec p,L}(Q_{\rm max})  -   2\,(d_{[\vec p],L}-1), }
then $\del_{\vec p,L}$ is the integer number  measuring quantum effects.
The explicit calculation employing the well--known   algorithm ${\rm u}(3)\to {\rm so}(3)$ reduction\cite{BohrMI}  gives
\begg{ \label{delF}   \del_{\vec p,L}=  \del + M_1\,t_1 + M_2\,t_2, \\
       \del = m_1 + m_2+ m_L - 2\,[(m_1+m_2 )\,M_L+m_1\,m_2] + 4\,m_1\,m_2\,m_L,\\
        \label{ti} t_i = \beg{cases}{ 1 & \text{if~}  L \le p_{i}-p_{i+1}, \\ 0 & \text{else},} \\
% \intertext{where}
\label{Mm} M_i = m_i +m_L - 2\,m_i\,m_L,\qquad
         m_i= {\rm Mod}_2(p_i-p_{i+1}),\qquad  m_L= {\rm Mod}_2(L).        }
Since $\del \in\{0,1\}$, $\del_{\vec p,L}$ is equal to $0,1,2$ or $3$.
The  Bohr Somerfield   quantization of momentum $Q$ is based  on the quantization of
the function $\c V_{\vec p,L}$ given by the integral in  eq.(\ref{volQ}).
It  takes the following form:
\begg{  \label{spQ}  \c V_{\vec p,L}(Q_{L,k}) = 2\,(k_u-1), \qquad
    k_u=k+u, \qquad  k\sS=1,\,2,\ldots, d_{[\vec p],L}, }
where $0\le u\le \del_{\vec p,L}$.
If $\lam=p_1-p_2$ or $\mu=p_2-p_3$ vanish,  then $\del_{\vec p,L}\equiv 0$ and $u \equiv 0$.
Here $u$ is unknown parameter which has to be fixed   as a function of $\lam,\mu,L$
by applying some  additional rules.
In the case $\lam\sS=\mu$ the parameter $u$ is found  from the symmetry.
Namely,
\llist{
\itt{a}
$q\mapsto \c V_{\vec p,L}(Q_{\rm max})-q=q\co$ is the reflection
of $M_{[\vec p],L}=\{ q, 0\le q  \le \c V_{\vec p,L}(Q_{\rm max}) \} $,
\itt{b}
 $ \tsf{D}_{\vec{p}}\ni (L,Q)  \mapsto \c V_{\vec p\cov ,L}(-Q) =  {\c V_{\vec p ,L}}\co (Q)\in M_{[\vec p],L} $ where
 $\vec p\cov=(-p_3,-p_2,-p_1)$,
 \itt{c} $g_u: k\mapsto q=2\,(k_u-1)$,  $k\cov \sS=d_{\vec p,L}+1 -k\Rightarrow
  {g_u}\ord{-1}\circ (g_u)\co (k)= k\cov+ (4\,u-\del_{\vec p,L})/2$.  }
 If   $\lam\sS=\mu$ then  $Q_{L,k}\mapsto -Q_{L,k}=Q_{L,k\cov}$
 is the rule of the physics symmetry  of the spectrum
 induced from the Poisson automorphism $\tau\!: (Q_{ab},L_c) \to (-Q_{ab},L_c)$. Thus,
the operation $q\to q\co$ is the physics  automorphism of $M_{\vec p,L}$ and
 the symmetry holds if  the term  $4\,u-\del_{\vec p,L}$ in the right hand side of point $(c)$ vanishes.

In order to exhibit the role of parameter $u$  we want to present pair of   solutions
violating the rule $u\sS=\del_{\vec p,L}/4$, i.e. which can be
considered if  $\lam\neq \mu$, only.  Let  $u(\vec p,L)\sS= u_s(\del_{\vec p,L})$ where
\begg{ \label{toUee}  u_s(0)=0, \qquad  u_s(1)= \tfrac 1 4\,(1+s), \qquad
         u_s(2)=u_s(3) = \tfrac 1 4\,(3+s),\qquad |s|\le 1.   }
If $|s|\sS<1$ then $0\le k_u \le \c V_{\vec p,L}$, thus
all  $(L,Q_{L,k_{u_s}})$  are points of the classic domain of coordinates $(L,Q)$.

The numerical results are  studied on the figure Fig.2 for multiplet  $[\vec p]\sS=[60,20,0]$.
The  left graphic presents $(L,Q)$ spectrum for  $s\sS=1$ while right one for $s\sS=-1$.
This parameter affects  the values $Q$ for $L$--odd states only.

The angular momentum states $L=1,2,\ldots,\lam+\mu$ form two sequences of bands
which are regarded by drawing the  solid and dashed  lines.
The lowest solid line joins the states forming $\tsf{P}_-$ ellipsoids.
Since  $\lam-\mu=p_1+p_2-2\,p_3>0$,
on the right picture ($s=-1$) the gap  $\Del L$ between band  states  is equal to one.
On the left one,  $L$ odd and $L$ even states  decouple into two bands, so $\Del L\sS=2$.

The highest dashed line selects states  of $\tsf{P}_+$ ellipsoids. For $s\sS=1$, $\Del L$ is equal to  $1$
while  $\Del L\sS=2$ when $s\sS=-1$.
The lowest dashed line selects  $\tsf{S}_3$ ellipsoids while the highest solid one joins $\tsf{S}_1$ states.
For these two bands  $\Del L\sS=2$.

Let $\Pi_s$  be    the sequences  of points:
$P_k \sS=(L_k,Q_k) = (4\,k, Q_{4\,k,k+1})$ where $4\,k\le {\rm min}(\lam,\mu)$ if  $s\sS=1$
and  $P_k \sS=(\bar L_k,Q_k) = (4\,k-1, Q_{4\,k-1,k})$ where $4\,k-1\le {\rm min}(\lam,\mu)$ if $s\sS=-1$.
The numerical results lead to  the   degeneracy $Q_1\sS=Q_2\ldots=\bar Q_\Pi(\vec p)$ where
$\bar Q_\Pi=\bar Q_{\sign(\lam-\mu)}({\rm min}\,(\lam,\mu))$ (see, eq.(\ref{QintL})); hence, if $\lam\le \mu$, then
\beqm{ Q_{\Pi}=\tfrac 1 {27}\,(p_1 - p_2)^{-2}\times
 [2\,(p_1 + p_2)^3 + 3\,(p_1 - 5\,p_2)\,(5\,p_1 - p_2)\,p_3 + 24\,(p_1 + p_2)\,{p_3}{^2} - 16\,{p_3}{^3}].
 \nonu }
else $Q_{\Pi} (\vec p)\sS=Q_{\Pi}(p_3,p_2,p_1)$.

For a sequence of states $\Pi$, $Q_\Pi(60,20,0)\approx 23.85$.
The approach does not predict the quantization of  states  $L\sS=0$.
Since  $d_{[60,20,0],0}\sS=1$, it is rather obvious that for singlet $L\sS=0$: $Q\sS=\bar Q_{\Pi} (\vec p)$.
On both the graphics  the states of $\Pi$ sequence  are joined by the dashed  lines including this singlet.

The formulas (\ref{toUee}) restore $(L,Q)$ structure in the cases  when $\lam$ and $\mu$ are  even numbers.
In order to  obtain  the similar families  of bands in the cases when
$\lam$ or/and $\mu$ are odd numbers, the function $u_s$ has to be modified $u_s\to \bar u_s$.
Using the same notation as in the formulas (\ref{ti},\ref{Mm})  one finds
\begg{   \bar u_s(\vec p,L) = u_s(\del_{\vec p,L}) - \tfrac 1 2\,\Del_s(\vec p,L), \\
        \Del_1(\vec p,L)=  M_2\,[m_1\,m_L+ m_2\,(1-m_L- m_1\,m_L)],  \\
        \Del_{-1} (\vec p,L)=  - m_2\,M_1\,[m_1 - M_2\,(1-m_L)   + m_L -2\,m_1\,M_L].    }
%L
%\[ \hskip -50mm \begin{comment}
   % \begin{comment}
\begin{figure}[!t]
  \centering    \includegraphics[scale=.8]{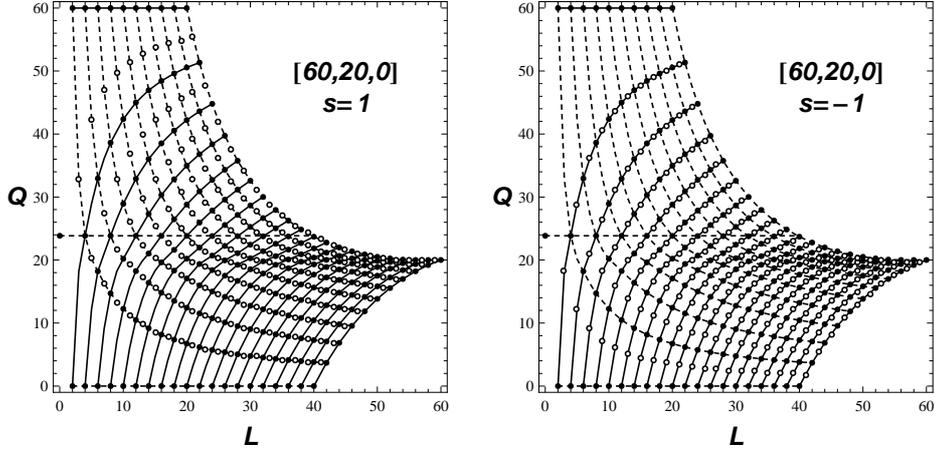}
  \caption{Bohr Somerfield's quantization of $(L,Q)$ momenta }
\end{figure}
    % \end{comment}
%
{  \Large \bf \vskip 10mm \noi  Appendix: A. {\rm SO}(2)  reduced Poisson bracket \\[5mm] }
\indent The formulas (\ref{gamma},\ref{bast},\ref{bast1}) determining the bracket $\{.,.\}_{\ast}$ are found
considering the following relations
\begg{ d\,[ai] = \beg{cases}{  \sum_{bc j }\,\eps_{abc}\,\,[bi]\,\bar \gam_{c j}(\vec{\tsf L},\tsf{L}_3)\,d\tsf{L}_j,  \\
              \sum_{bc}\,\gam_{abc}\,[bi]\,dL_c,} \\
  \label{togab} \bar \gam_{ci}(\vec x,u) =  \hph{-}\del_{cz}\,f_{1}(\vec x,u)\,\sum_a \eps_{i m(a)  3}\,x_a
      +  u^{-1}\,\eps_{m(c)i3},\\
\gam_{i j}(\vec x,u)=  -\del_{i3}\,\,f_{1}(\vec x,u)\,\Sig_{a}\,\eps_{jm(a)3}\,x_{a} + u^{-1}\,\eps_{ij3}, \\
  \gam_{abc}\sS =  \sum_{d }\eps_{abd}\, \gam_{dc} = \sum_{id } \eps_{abd}\, \bar \gam_{di}\,[ci],  }
where $[ai]\sS= [{\rm sign}\,(L_3) \,\vec{\tsf L} \,]_{m(a) i}$,
$a,b,\ldots =x,y,z$ and $i,j,\ldots =1,2,3$, $m(x)\sS=1,\,m(y)\sS=2,m(z)\sS=3$ and
$\eps_{abc}$ is the antisymmetric tensor that $\eps_{xyz}\sS=1$ and
$  f_{k}(\vec x,u)= u^{-k}\,(u + x_z )^{-1}$.
Using the above notation and applying Leibnitz rule to formula eq.(\ref{bast}) one finds
\begg{ \label{toqci}   \sum_{i}   \{ q^u_a,[c i] \}\,[d i] =
   \Sig_{bd e} \eps_{a e b}\,\gam_{c d e}(\vec l,l_3)\,q^u_b =
    \Sig_b\,\Gam^{1}_{ab\,cd}(\vec l,l_3)\,q^u_b, \\
  \sum_{i j}   \{ [a i],[c i]\}\,[b i]\,[d j] = \sum_{}  \gam_{abe}(\vec l,l_3)\,
         \gam_{cd f} (\vec l,l_3)\,\eps_{efg}\,l_g = \Gam^{2}_{ab\,cd}(\vec l,l_3),  \\
  \Gam_{ij }(\vec x,\vec y,\vec l,l_3)= \sum_{kl} x_k\,y_l\,
  \sum_{a b c d}  \,(\Gam^{1} _{ab\, cd}-\Gam^{1}_{cd\, ab}+
             \Gam^{2}_{ab\,cd})(\vec l,l_3)\,[a i]\,[ b k]\,[c j] [d\, l] }
The  explicit calculations give
\begg{  \label{G1} \Gam^{1}_{ab\,cd}(\vec x,u) = f_{2}(\vec x,u)\,\sum_{ef}\eps_{abe}\,\eps_{cdf}\,
     [ \varepsilon_f\,(1-\del_{e z})\,x_{e}\,x_{3-f} + P_{ef}(\vec x,u) ],     \\
    \label{G2}  \Gam^{2}_{ab\,cd}(\vec x,u)= f_{1}(\vec x,u)\,
     \sum_{efg}\eps_{abe}\,\eps_{cdf}\,\eps_{efg}\,(\eps\,x_g +\del_{g3}\,x), \\
    \label{Pef}   P_{ef}(\vec x,u) = \sum_g \eps_{efg}\,(u + \del_{e z}\,x_e)\,  (x_g + \del_{g z}\,u),  }
where  $\varepsilon_x\sS=-\varepsilon_y\sS=1,\,\varepsilon_z\sS=0$ and $\vec x=(x_x,x_y,x_z)$.
Comparing these formulas with eq.(\ref{bast1}) one finds
$ \Gam_{ij}(\vec x,\vec y,l_3)= \Gam_{ij}(\vec x,\vec y,\vec l,l_3)$; hence,
$\gam^{kl}_{ij}= {l_3}\,\sum_{a b c d}  \,(\Gam^{1} _{ab\, cd}-\Gam^{1}_{cd\, ab}+
             \Gam^{2}_{ab\,cd})(\vec l,l_3)\,[a i]\,[ b k]\,[c j] \\ \times {[d\, l]}$
which  with help of equations (\ref{G1},\ref{G2},\ref{Pef}) restate the formula eq.(\ref{gamma}).

In order to close the system of Poisson relations related
with coordinates system $(\bo b^1,\bo b^2,\vec{\tsf L})$  where
$L(A(\bo b^1,\bo b^2))=(0,0,L_3\sS=\pm |\vec{\tsf{L}}| )$   let us write,
\beqm{  \{ \tsf{L}_a  ,b_{iun} \} = h_{aiun}=   \,\psi_a(\vec {\tsf{L}},L_3)\,\sum_{j}\,g_{i j}\,b_{jun}.    }
Since $b_{iun}\sS=\Sig_{b} [bi]\,a_{bun}$; hence,
\begg{ h_{aiun}= \Sig_{b c }\, \eps_{abc}\,([b i]\,a_{cun} +  \Sig_{ d e }
   \gam_{de  b} \,[e i]\,a_{dun}\,\tsf{L}_c )
  = \Sig_{jk}\, [aj ]\,  ( \eps_{jik} +  \Sig_{ l   }\, \eps_{j l 3}\, \gam_{k i l} \,L_3) \,b_{kun},    }
\beqm{  \label{topsi}    \vec \psi (\vec x, y )   =   {\rm sign}(y)\,\frac{ \vec x   +   \vec e_3 \,y }{x_3 + y}, \qquad
           g_{ij}= \eps_{ij3}.     }
From  eq.(\ref{Lxyz}):
$\psi_a(\vec {\tsf L}, L_3) =  \pl_{L_3}  \tsf{L}_\a =   \{\vp_L, \tsf{L}_a \}$
where  we assumed $\{\vp_L, p_L \}= 1,\,p_L\sS\equiv |L_3|$. If
\beqm{ \label{L123} \c L_1 \pm i\,\c L_2= e^{i\,\vp_K}\,\sqrt{\smash[b]{ p_K\,(2\,p_L - p_K)} } , \qquad  \c L_3= p_L  - p_K, \qquad
         \{\vp_X,p_Y \} = \del_{X Y},   }
where $X,Y\sS=M,L,K$ then   we have too: $\{\vp_L,\c L_{\a}\} = \pl_{L_3}\, \c L_3=  \psi_{\a}(\vec{\c L},L_3)$ which
restate the  rules in eq.(\ref{RB1}).
%verified applying eqs.(\ref{togab},\ref{toqci}).
%

%verified applying eqs.(\ref{togab},\ref{toqci}).
%
 % \end{document}
\def\tspr#1{\textsuperscript{#1}}

\end{document}